\documentclass[twocolumn]{aastex63}

\usepackage{sidecap}
\usepackage{wrapfig}
\usepackage{float}
\usepackage{xcolor}

\received{}
\revised{}
\accepted{}
\submitjournal{AJ}

\shorttitle{Helium in sub-Neptunes}
\shortauthors{Kasper et al.}
\graphicspath{{./}{figures/}}

\begin{document}

\title{Non-detection of Helium in the Upper Atmospheres of Three Sub-Neptune Exoplanets}

\correspondingauthor{J.\ L.\ Bean}
\email{jbean@astro.uchicago.edu}

\author[0000-0003-0534-6388]{David Kasper}
\affil{Department of Astronomy \& Astrophysics, University of Chicago, 5640 South Ellis Avenue, Chicago, IL 60637, USA}

\author[0000-0003-4733-6532]{Jacob L.\ Bean}
\affil{Department of Astronomy \& Astrophysics, University of Chicago, 5640 South Ellis Avenue, Chicago, IL 60637, USA}

\author[0000-0002-9584-6476]{Antonija Oklop{\v{c}}i{\'c}}
\affiliation{Center for Astrophysics | Harvard \& Smithsonian, Cambridge, MA 02138, USA}
\affil{NHFP Sagan Fellow}

\author[0000-0003-0217-3880]{Isaac Malsky}
\affil{Department of Astronomy \& Astrophysics, University of Michigan, Ann Arbor, MI, 48109, USA}

\author[0000-0002-1337-9051]{Eliza M.-R.\ Kempton}
\affil{Department of Astronomy, University of Maryland, College Park, MD 20742, USA}

\author[0000-0002-0875-8401]{Jean-Michel D\'esert}
\affiliation{Anton Pannekoek Institute for Astronomy, University of Amsterdam, 1090 GE Amsterdam, Netherlands}

\author[0000-0003-0638-3455]{Leslie A. Rogers}
\affil{Department of Astronomy \& Astrophysics, University of Chicago, 5640 South Ellis Avenue, Chicago, IL 60637, USA}

\author[0000-0003-4241-7413]{Megan Mansfield}
\affil{Department of Geophysical Sciences, University of Chicago, 5734 South Ellis Avenue, Chicago, IL 60637, USA}



\begin{abstract}
We present a search for helium in the upper atmospheres of three sub-Neptune size planets to investigate the origins of these ubiquitous objects. The detection of helium for a low density planet would be strong evidence for the presence of a primary atmosphere accreted from the protoplanetary nebula because large amounts of helium are not expected in the secondary atmospheres of rocky planets. We used Keck+NIRSPEC to obtain high-resolution transit spectroscopy of the planets GJ\,1214b, GJ\,9827d, and HD\,97658b around the 10,833\,\AA\ He triplet feature. We did not detect helium absorption for any of the planets despite achieving a high level of sensitivity. We used the non-detections to set limits on the planets' thermosphere temperatures and atmospheric loss rates by comparing grids of 1D models to the data. We also performed coupled interior structure and atmospheric loss calculations, which suggest that the bulk atmospheres (winds) of the planets would be at most modestly enhanced (depleted) in helium relative to their primordial composition. Our lack of detections of the helium triplet for GJ\,1214b and GJ\,9827d are highly inconsistent with the predictions of models for the present day mass loss on these planets. Higher signal-to-noise data would be needed to detect the helium feature predicted for HD\,97658b. We identify uncertainties in the EUV fluxes of the host stars and the lack of detailed mass loss models specifically for cool and metal-enhanced atmospheres as the main limitations to the interpretation of our results. Ultimately, our results suggest that the upper atmospheres of sub-Neptune planets are fundamentally different than those of gas giant planets.
\end{abstract}



\section{Introduction} \label{sec:intro}

\begin{SCfigure*}[][!t]
\includegraphics[width=1.25\linewidth]{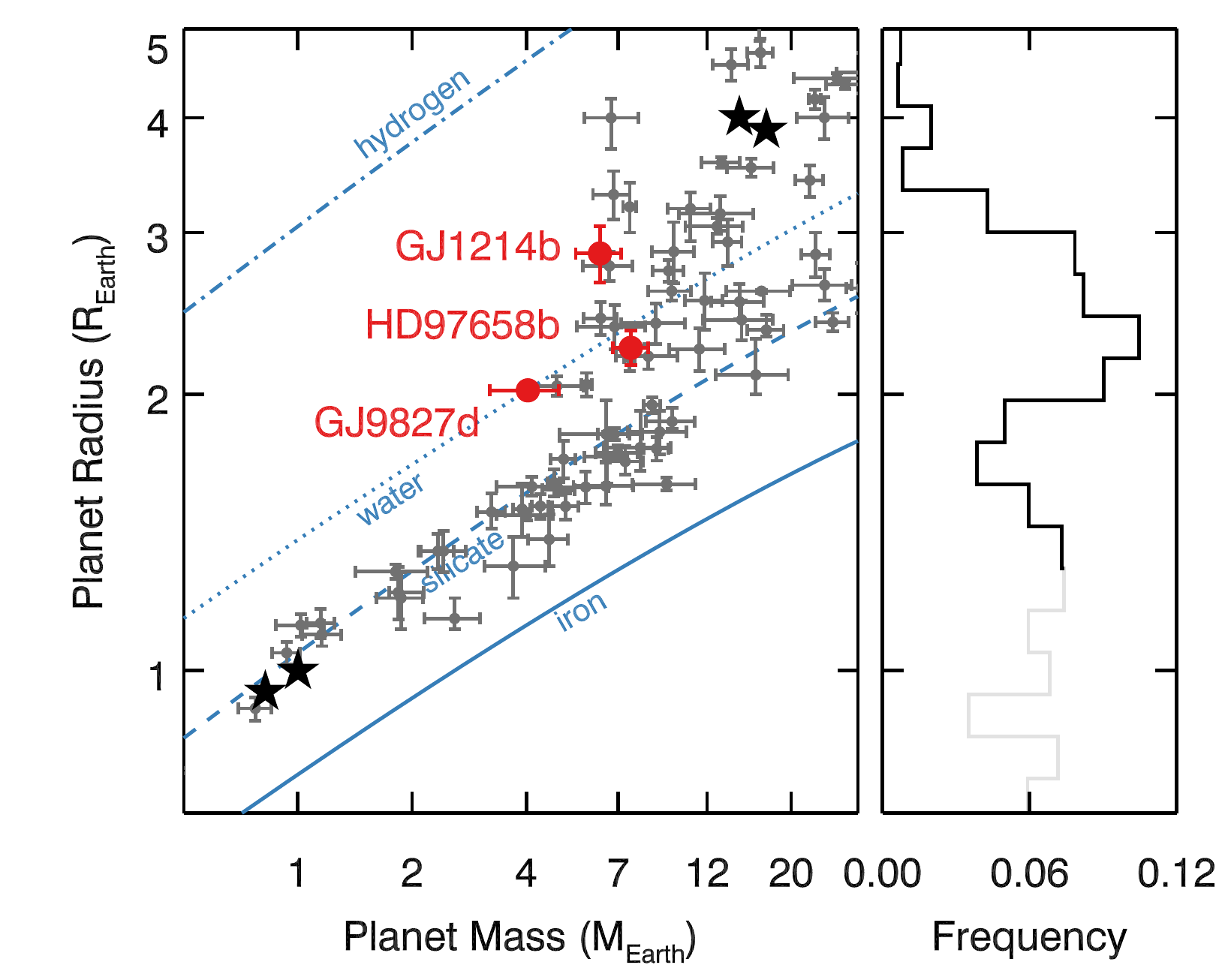}
\caption{\textbf{The targets of our project in the context of the broader population of close-in exoplanets.} \textit{Left:} Mass-radius diagram showing solar system planets (black stars), exoplanets with precisely measured parameters \citep[grey circles, data from][]{otegi20}, and our targets (red circles). The lines are theoretical models for pure compositions of iron, silicate, water, and hydrogen \citep{seager07}. \textit{Right:} Frequency of close-in planets as a function of planet size from \citet{fulton18}. The grey part of the line indicates the regions of parameter space with poor constraints.}
\label{fig:sample}
\end{SCfigure*}

The basic nature of the atmospheres of sub-Neptune size planets, whether they are primary or secondary, is currently a major question in the field of exoplanets. The bimodality of the small planet radius distribution in \textit{Kepler} data and its dependence on stellar irradiation is strong evidence that these planets have hydrogen-dominated atmospheres accreted from the primordial nebula \citep[e.g.,][]{fulton17, fulton18, vaneylen18, owen17}. However, it has been difficult to test this hypothesis using direct observations of the atmospheres themselves due to impact of aerosols on transmission spectra and the concomitant challenge of determining precise chemical abundances from low information content data \citep[e.g.,][]{kreidberg14, knutson14, benneke19, tsiaras19}. Furthermore, even if we could determine that these planets' atmospheres were hydrogen-dominated this would still not be definitive proof that they were accreted from the primordial nebula because hydrogen can be outgassed from the interiors of rocky planets \citep{elkins-tanton08, rogers10b, rogers11, chachan18, kite20}.

In this paper we pursue a new way to investigate the atmospheric compositions of the mysterious sub-Neptune size exoplanets. We aimed to detect helium in these planets' upper atmospheres (i.e., the thermospheres and exospheres) using transmission spectroscopy. The thermosphere of a planet is the outermost bound atmospheric layer, above which is the exosphere and the transition to interplanetary space. The upper atmospheres of close-in exoplanets are expected to be extended due to substantial ongoing atmospheric escape driven by absorption of UV flux from their host stars. The advantage of targeting the thermospheres and exospheres for sub-Neptune size planets is that they should extend to altitudes well above the aerosol layers that blocks transmission spectroscopy observations of their bulk atmospheres. That is, we shouldn't be blocked from viewing the upper atmospheres by the aerosols that have caused previous transmission spectroscopy observations to yield flat spectra. If helium is present in the atmospheres of these planets then it should be abundant in the thermospheres and exospheres because it is a relatively light species.

Our observations target the 10,833\,\AA\ He triplet feature, which has recently emerged as a way to probe the escaping atmospheres of larger planets \citep{spake18, nortmann18, allart18, allart19, mansfield18, salz18, kreidberg18, alonsofloriano19, ninan20, kirk20, gaidos20, palle20, vissa20, dossantos20}. The He triplet feature arises from a metastable transition. It has a strong absorption cross section that compensates for the low density of the upper atmosphere and allows detection even at very low partial pressures.

Our idea is similar to searching for hydrogen gas in the upper atmosphere using Lyman\,$\alpha$ \citep{bourrier17}, but with the advantage that a detection of helium would have a less ambiguous interpretation. The direct detection of helium in the atmosphere of a planet would be the smoking gun evidence for the presence of a primary atmosphere accreted from the primordial nebula. This is because no other formation mechanism (e.g., sublimation of ices or outgassing of rocky material) can give rise to the presence of large quantities of helium in a planetary atmosphere \citep{elkins-tanton08}. Helium is thus actually a cleaner diagnostic of formation history than the now standard transmission spectroscopy scale height test \citep{miller-ricci09}. Furthermore, the helium infrared triplet that we target doesn't suffer from absorption by the interstellar medium like Lyman\,$\alpha$.

We present Keck+NIRSPEC transit spectroscopy observations of the sub-Neptune size exoplanets GJ\,1214b \citep{charbonneau09}, HD\,97658b \citep{howard11,dragomir13}, and GJ\,9827d \citep{niraula17,rodriguez18} to search for the 10,833\,\AA\ He triplet feature. Of these, only GJ\,1214b has a previously published result on the He triplet. Using low resolution (R\,$\approx$\,500) archival IRTF/SpeX observations, \citet{crossfield19} were able to place a modest upper limit of 2.1\,$R_{p}$ at 95\% confidence for absorption in the line core. Our observations reach much higher sensitivity for this planet due to the higher spectral resolution and signal-to-noise, and the lower levels of systematic noise in our data.

The adopted properties of our targets are given in Table~\ref{tab:modeling_parameters} and placed in the context of the population of sub-Neptune size planets in Figure~\ref{fig:sample}. These planets are archetypal intermediate-size planets as they sit squarely in the middle of the degenerate mass-radius space where there are multiple plausible models for their internal structure \citep[e.g.,][]{adams08, rogers10a}. Our targets were chosen for having the highest expected signal-to-noise for planets in this part of parameter space (formally, planets having $M_{p}\,<\,10\,M_{\oplus}$ and $R_{p}\,>\,2\,R_{\oplus}$) and being observable from Mauna Kea.

GJ\,1214b in particular has been the subject of intense scrutiny that has revealed an astonishingly high and opaque aerosol layer that is challenging to explain \citep{kreidberg14, morley15, gao18, adams19}. HD\,97658b has also been observed to have a featureless transmission spectrum \citep{knutson14, guo20}. These two planets remain the best targets for transmission spectroscopy observations in this part of the parameter space, and GJ\,9827d is not far behind. \textit{TESS} is not expected to find better planets for this study \citep{louie18, kempton18}.

\begin{deluxetable}{llc}
\tablecolumns{3}
\tablewidth{0pc}
\tablecaption{\label{tab:modeling_parameters} Adopted physical properties of the targets \tablenotemark{a}}
\tablehead{
 \colhead{Planet} & \colhead{Parameter} &
 \colhead{Value}
}
\startdata
{\bf GJ\,1214b} & & \\
 & planet mass ($\text{M}_{\oplus}$) & 6.26\\
 & planet radius ($\text{R}_{\oplus}$) & 2.85 \\
 & stellar mass ($\text{M}_{\odot}$) & 0.15 \\
 & stellar radius ($\text{R}_{\odot}$) & 0.22 \\
 & semi-major axis (AU) & 0.01411 \\
 & impact parameter &  0.38 \\
{\bf HD\,97658b} &  & \\
 & planet mass ($\text{M}_{\oplus}$) & 7.82 \\
 & planet radius ($\text{R}_{\oplus}$) & 2.24 \\
 & stellar mass ($\text{M}_{\odot}$) &  0.89 \\
 & stellar radius ($\text{R}_{\odot}$) & 0.74 \\
 & semi-major axis (AU) & 0.0800 \\
 & impact parameter &  0.35 \\
{\bf GJ\,9827d} & & \\
 & planet mass ($\text{M}_{\oplus}$) & 4.04 \\
 & planet radius ($\text{R}_{\oplus}$) & 2.02 \\
 & stellar mass ($\text{M}_{\odot}$) &  0.61 \\
 & stellar radius ($\text{R}_{\odot}$) & 0.60 \\
 & semi-major axis (AU) & 0.05591 \\
 & impact parameter &  0.89 \\
\enddata
\tablenotetext{a}{Data taken from the TEPCat: \url{https://www.astro.keele.ac.uk/jkt/tepcat/}, \citet{southworth11}}
\end{deluxetable}

The planets we observed differ in terms of their host stars and system multiplicity, thus providing an interesting comparison set. GJ\,1214b orbits an M dwarf, while HD\,97658b and GJ\,9827d orbit early and late K dwarfs, respectively. Looking at planets orbiting different stellar types is important for this study because both photoevaporative mass loss and the population of the metastable level that gives rise to the 10,833\,\AA\ He triplet feature depend on the high-energy irradiation received from the host star. \citet{oklopcic19} have suggested that K stars in particular have the ideal balance between extreme-ultraviolet (EUV) and mid-ultraviolet (mid-UV) flux levels. Beyond the stellar hosts, no other planets are known in the GJ\,1214 \citep{gillon14} and HD\,97658 systems. However, GJ\,9827d is part of a system with two other transiting planets on interior orbits. The density of the closest-in planet (planet b) is consistent with being rocky \citep{prieto18,teske18,rice19}, which is highly suggestive of photoevaporative mass loss from initially volatile-rich worlds. GJ\,9827c is between planets b and d, but its mass is poorly constrained.

The paper is laid out as follows. We describe our observations and data analysis in \S\ref{sec:obs} and \S\ref{sec:analysis}. Since we do not detect the He triplet feature for any of the planets we perform a suite of atmosphere (\S\ref{sec:model}) and interior structure (\S\ref{sec:interior}) model calculations to interpret the upper limits from our data. We discuss the implications of our results in \S\ref{sec:discussion}, and we conclude in \S\ref{sec:summary} with a summary.

\begin{deluxetable*}{llccccccccclc}
\tabletypesize{\scriptsize}
\tablecolumns{13}
\tablewidth{0pc}
\tablecaption{\label{tab:observation_log} Log of observations}
\tablehead{
 \colhead{Target} & \colhead{UT Date} & \colhead{Exposure Time (s)} & \colhead{\#} &  \colhead{Airmass} & \colhead{Conditions} & \colhead{Seeing}
}
\startdata
{\bf GJ\,1214b} & 2019 August 21 05:42 $\rightarrow$ 09:08 & 140,160 & 73 & 1.03 $\rightarrow$ 1.12 $\rightarrow$ 1.64 & clear & 0.6\arcsec\\ 
{\bf GJ\,9827d} & 2019 October 26 04:59 $\rightarrow$ 10:02 & 150, 250, 500 & 77 &  1.37 $\rightarrow$ 1.07 $\rightarrow$ 1.35 & cloudy in first 1/3 & 0.8\arcsec\\
            & 2019 November 26 05:07 $\rightarrow$ 08:48 & 100 & 94 & 1.08 $\rightarrow$ 1.16 $\rightarrow$ 1.66 & clear & 0.9\arcsec\\
{\bf HD\,97658b} & 2020 February 06 08:18 $\rightarrow$ 08:45, 10:22 $\rightarrow$ 11:05 & 60 & 46 & 1.91 $\rightarrow$ 1.66, 1.16 $\rightarrow$ 1.07 & variable heavy clouds &  1.2\arcsec\\
\enddata
\end{deluxetable*}		

\section{Observations} \label{sec:obs}
We observed transits of our targets using the NIRSPEC instrument \citep{mclean98} on the Keck II telescope on Mauna Kea, Hawai'i. An observing log is given in Table~\ref{tab:observation_log}. Our program was conducted following the NIRSPEC hardware upgrades described by \citet{martin18}. We used the 0.432\arcsec $\times$ 12\arcsec\, slit, which delivers a nominal resolving power $R\,\approx$\,25,000. We used the NIRSPEC-1 (Y-band) filter setting without the optional THIN blocking filter to avoid fringing effects at the recommendation of the Keck instrument scientist. We took observations in a standard ABBA nodding sequence with a throw of 6\arcsec. Dark, flat, and arc-lamp calibration frames were taken at the beginning and end of the observation periods.

We observed one transit each of GJ\,1214b and HD\,97658b, and two transits of GJ\,9827d. One additional planned transit observation of HD\,97658b in April 2019 was lost due to a telescope shutdown caused by the COVID-19 pandemic. We observed our targets continuously starting well before transit ingress and ending well after transit egress except for minor interruptions due to telescope and software glitches and major interruptions due to bad weather during the HD\,97658b transit. For the HD\,97658b transit, the dome had to be closed for 97\,minutes during ingress and then closed for the night just after mid-transit. Nevertheless, we obtained a series of spectra both before and during transit for HD\,97658b, and the data quality for this target is useful due to the very bright host star ($m_{J}\,=\,6.2$) and the large collecting area of the 10\,m Keck telescope. The weather conditions were generally clear for the other transits. All observations were taken at airmass values less than 2.0, with most at less than 1.5.

\section{Data Analysis} \label{sec:analysis}
We analyzed the data for this project using custom routines that have been previously used to reduce ground-based low- and high-resolution spectroscopy data for atmospheric studies and radial velocity measurements \citep{bean10b, bean10a, bean11}. The routines were originally written in \texttt{IDL} and were ported to \texttt{Python}, version 3.7, for this project. In addition to our own data, we also reduced the Keck+NIRSPEC WASP-107b data described in \citet{kirk20} as a check of our pipeline. This planet was the first to show the helium feature, and the feature has been detected for this planet using \textit{HST}+WFC3, CARMENES, and NIRSPEC, with consistent results from all the instruments \citep{spake18, allart19}. The helium feature appears in two of the NIRSPEC echelle orders (70 and 71), and we limited our analysis to just these orders. The scripts used for this paper are available from the authors upon request.

\subsection{Data reduction}

\textit{Processing the calibration frames} -- Our NIRSPEC data reduction starts with creating a master dark image for the flat field exposures by averaging exposures taken with a cold blank in the filter wheel blocking the light path. The dark frames were obtained using the same exposure time as the flat fields. We then subtracted this master dark image from the flat field exposures to remove the signals from the bias (pedestal) and dark currents. The flat field exposures were then averaged to create a single master flat field image. We removed the spectroscopic signature from the master flat for each order considered by averaging the rows in the cross dispersion direction, applying a median filter to smooth the data, fitting a high order polynomial, and then dividing out the fit. We then normalized the flat by dividing by the median of all the illuminated pixels in an order. In most of the averaging steps we used iterative outlier rejection with cutoffs ranging from three to five standard deviations. We created an initial bad pixel mask from the flat normalization process that was later used in the spectral extraction.

\textit{Processing the science frames} -- The science frames were first processed by subtracting the closest-in-time opposite nod pair frame taken with the same exposure length. This simultaneously removed the bias and dark currents, and the sky background (both continuum and emission line). Telluric emission lines can be a pernicious problem for data taken with fiber-fed spectrographs to search for the helium feature because the source and background are scrambled and the spatial information is lost \citep[e.g.,][]{ninan20,palle20}. Our slit-fed observations sacrifice stability (see later in this section) but make accurate sky subtraction straightforward. 

After subtracting the nod pairs we divided the science frames by a normalized flat field. Then we traced the orders and used an optimal extraction algorithm \citep{horne86} to extract 1D spectra from the images. Our optimal extraction algorithm uses a spatial profile weighting that is determined from the data and includes iterative identification and masking of bad pixels and cosmic rays. We used an aperture radius of 10 pixels around the center of the spectral trace for the extraction, and we verified that the results did not depend on our choice for this parameter.

\begin{figure}[t!]
\begin{center}
\includegraphics[width=\linewidth]{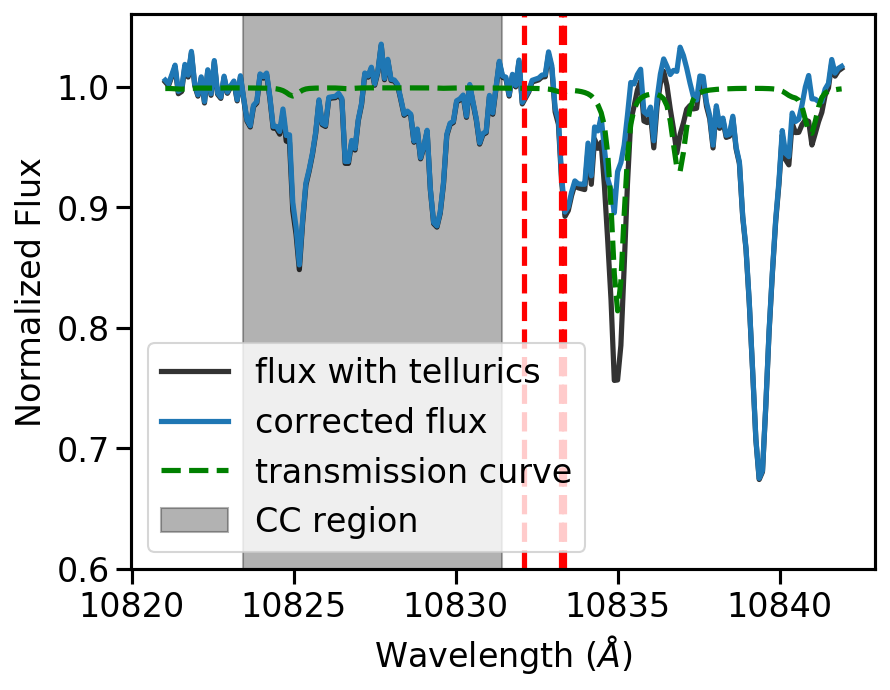}
\caption{\textbf{Example telluric correction around the 10,833\,\AA\ feature using Molecfit.} The continuum normalized GJ\,1214 spectrum is shown prior to (black) and post correction (blue) in the rest frame of the observatory. The Molecfit best-fit transmission curve -- the correction function for telluric absorption -- is shown in dashed green. The gray region is used to correct instrumental drift between frames and is notably free of significant telluric features that could compromise the cross-correlation technique. Red dashed vertical lines indicate where the line center positions of the helium triplet would be at mid-transit for planet b in this reference frame.} 
\label{fig:telluric_removal}
\end{center}
\end{figure}

\textit{Wavelength calibration} -- To provide an initial wavelength calibration of the data, we used a second order polynomial to fit the positions of emission lines bracketing the 10,833\,\AA\ feature along each order in nightly NeArXeKr arc-lamps exposures, with six lines in order 70 and five lines in order 71. We did this separately for the A and B nod positions but we found that this wasn't necessary because the slit is well aligned with the columns on the detector. As a reminder, even though these data are from ground-based observations, the wavelengths are measured in vacuo because the detector is in a cryo-vacuum dewar. Contrary to \citet{kirk20}, we did not find that our initial wavelength solution based on the arc lamp data had any obvious distortions. During this step we also confirmed that the data have a resolution R\,$\approx$\,25,000 by measuring the width of the arc lamp lines.

\begin{figure}[t!]
\begin{center}
\includegraphics[width=\linewidth]{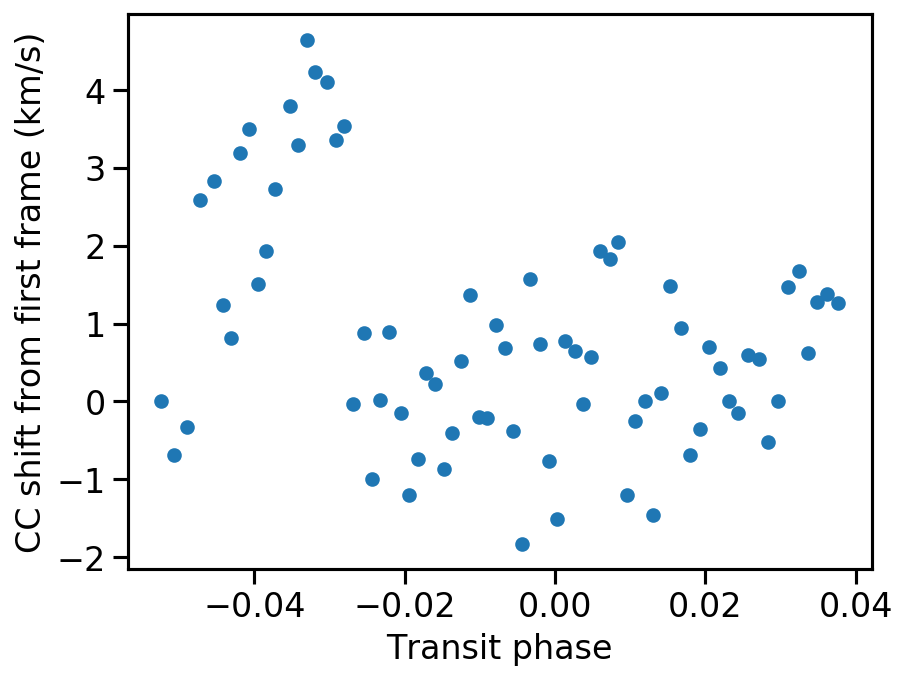}
\caption{\textbf{Drift correction for GJ\,1214b data.} The cross-correlated shift in km\,s$^{-1}$ for each frame from the first frame is plotted as a function of phase. Care was taken in picking the wavelength region to perform the cross-correlation on to be near the helium feature but avoid potential contamination from residual telluric features and planetary features.} 
\label{fig:drift_cor}
\end{center}
\end{figure}

\textit{Removing telluric absorption lines} -- Following \citet{kirk20}, we ran \texttt{Molecfit} \citep{smette15, kausch15} separately for the two orders in each frame to remove telluric absorption lines. We did this before applying barycentric corrections to the wavelength scale because the telluric features appear in the rest frame of the observatory. We found that by-hand identification of clean telluric feature regions was necessary to optimize the \texttt{Molecfit} model parameters. The exact number of clean telluric features varied by target. For example, we fit 13 features in order 70 and 8 features in order 71 in the GJ\,1214b dataset to optimize the model parameters.

We allowed the H$_{2}$O abundance to vary and assumed the nominal values of the CH$_{4}$ and CO$_{2}$ abundances. \texttt{Molecfit} divided out the best-fit telluric model determined for each order and frame. An example spectrum before and after telluric correction for GJ\,1214 is shown in Figure \ref{fig:telluric_removal}. We found, as expected, that the coincidental overlap of telluric absorption features (in the Earth frame) and the helium triplet feature (in the system frame) was different for each planet and observation due to differences in the systemic velocities and the motion of the observatory along the line of sight.  

\begin{deluxetable*}{llccc}
\tablecolumns{5}
\tablewidth{0pc}
\tablecaption{\label{tab:transit_param} Adopted ephemerides for the targeted planets}
\tablehead{
 \colhead{System} & \colhead{Parameter} &
 \multicolumn{3}{c}{Value}
}
\startdata
{\bf GJ\,1214b}  & & & & \\        
\citet{gillon14} & $P$ (d) & 1.58040417 & $\pm$ & 0.00000016 \\
  & T$_{c1}$ (BJD$_{TBD}$) & 2454980.74900 & $\pm$ & 0.00010 \\
\citet{berta12}   & Transit Duration (hr) & 0.87 & & \\
\citet{charbonneau09} & Systemic Velocity (km/s) & +21.1 & $\pm$ & 1 \\[3mm]
{\bf GJ\,9827d} & & & & \\
\citet{rice19}  & $P$ (d) &  6.20147 & $\pm$ & 0.00006 \\
 & T$_{c1}$ (BJD$_{TBD}$) & 2457740.9612 & $\pm$ & 0.0004 \\
                 & Transit Duration (hr)  & 1.22 & $\pm$ & 0.03 \\
\citet{sperauskas16} & Systemic Velocity (km/s) & +32.1 & $\pm$ &  0.7 \\[3mm]
{\bf HD\,97658b} & & & & \\
\citet{guo20} & $P$ (d) & 9.489295 & $\pm$ & 0.000005 \\
 &  T$_{c1}$ (BJD$_{TBD}$) & 2456361.8069 & $\pm$ & 0.0003 \\
\citet{VanGrootel2014} & Transit Duration (hr) & 2.85 & $\pm$ & 0.03 \\             
\citet{gaiadr2} & Systemic Velocity (km/s) & -1.579 & $\pm$ &  0.001 \\
\enddata
We utilize the \citet{kirk20} phase determinations to compare directly with their results
\end{deluxetable*}	

\begin{figure}[t!]
\begin{center}
\includegraphics[width=\linewidth]{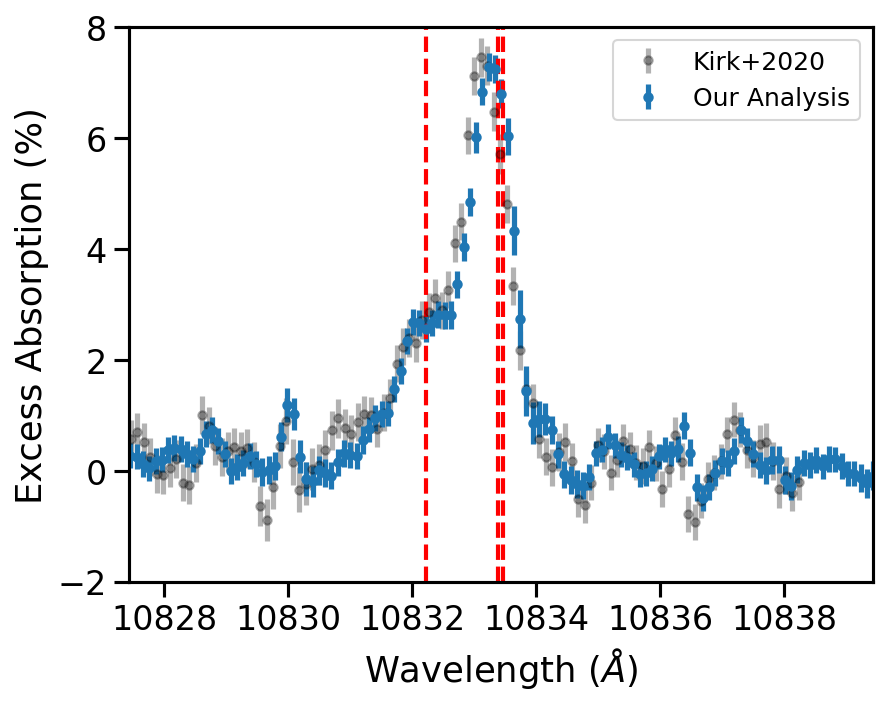}
\caption{\textbf{Excess absorption around the 10,833\,\AA\ feature for WASP-107b.} The \citet{kirk20} result for order 70 in the planet rest frame is plotted in gray. Our combined result for orders 70 and 71 is plotted in blue. The vertical red lines indicate the line centers for the helium triplet feature.} 
\label{fig:WASP107_analysis}
\end{center}
\end{figure}

\textit{Correcting for radial velocity shifts} -- After the telluric correction, we applied shifts to the wavelength scales of each frame to put them in the rest frames of the corresponding systems (i.e., accounting for the motion of the observatory relative to the barycenter of the systems). The adopted systemic velocities used for this correction are given in Table \ref{tab:transit_param}. Note that there is a sign error for the radial velocity of the GJ\,1214 system given by \citet{charbonneau09}. The correct systemic velocity should be +21.3\,km\,s$^{-1}$ (i.e., the star is moving away from the solar system, private communication from E.~Newton). Following this step we normalized the continuum in the data by fitting a line to relatively clean spectral regions within 15\,\AA\ to either side of the helium feature and then dividing it out.

We noticed that the time series of spectra for each observation do not match up perfectly after ostensibly putting them in the system rest frame, which we attribute to a combination of variations in the illumination of the entrance slit and instability of the spectrograph optics. We corrected for this effect by cross correlating the spectra against the first spectrum taken during each observation and then shifting the spectra according to the measured values. An example of the drift in the spectra during the GJ\,1214b observation is given in Figure \ref{fig:drift_cor}. The typical rms of the offset in the timeseries is 1.3\,km\,s$^{-1}$, which corresponds to 0.46\,pixels.

\begin{figure*}[t!]
\begin{center}
\includegraphics[width=\linewidth]{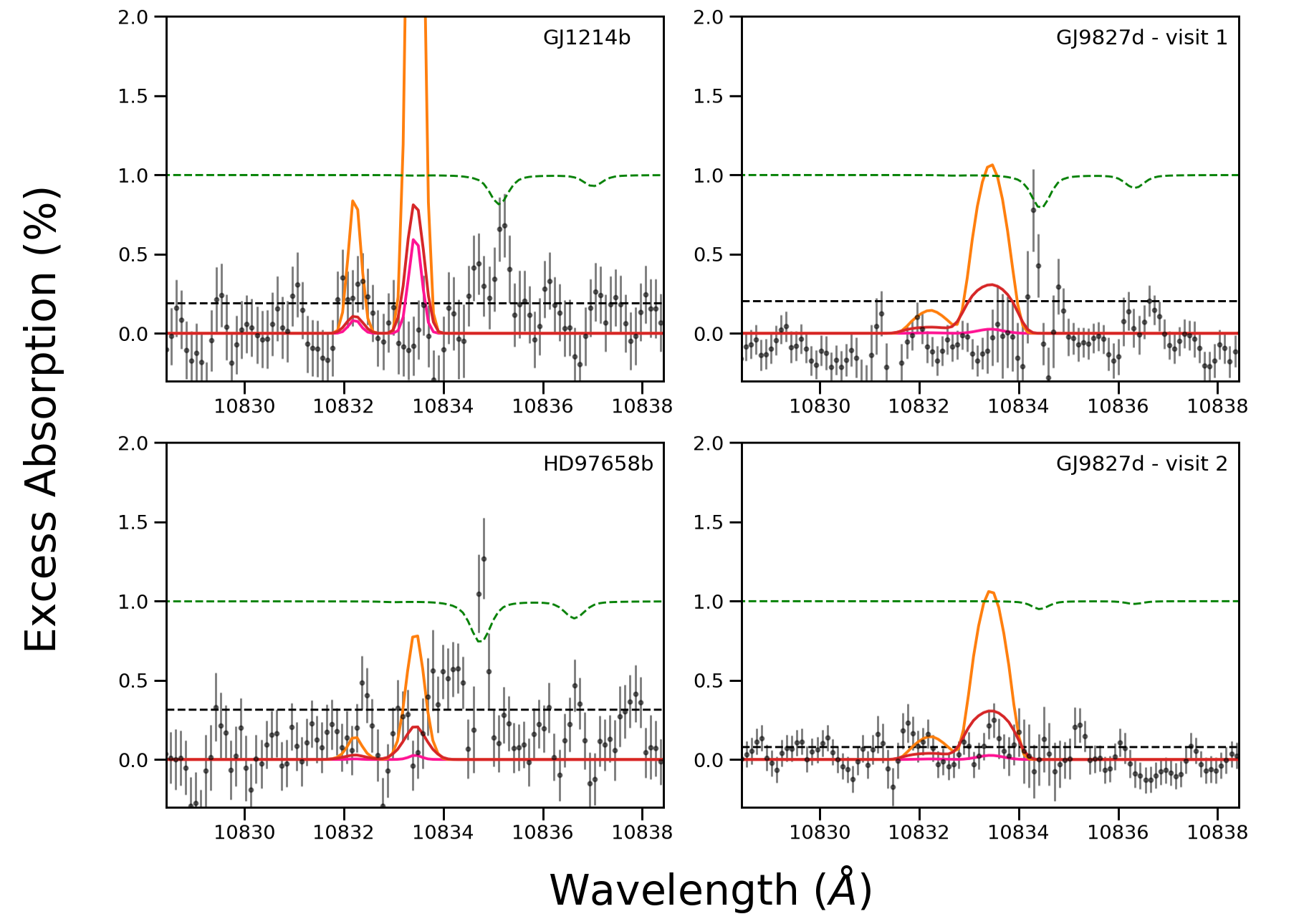}
\caption{\textbf{Measured transmission spectra for our targets compared to models.} The observed data are shown as grey points with errors. Representative models (see \S\ref{sec:model}) are overplotted as the orange, red, and magenta lines. The color coding corresponds to models represented by the positions of the circles in the parameter space maps in Figures~\ref{fig:GJ1214_models}, \ref{fig:GJ9827_1126_model_rejection}, and \ref{fig:HD97658_model_rejection}. The orange lines corresponds to the nominal models for each planet based on the calculations of \citet{salz16}. All example models shown have a solar helium abundance. The green dashed lines indicate the telluric transmission spectra shifted into the rest frame of each stellar system. The black dashed line indicates the standard deviation of the data around the helium feature.} 
\label{fig:spectra}
\end{center}
\end{figure*}

\subsection{Creation of the transmission spectrum}
Our final analysis step was to create transmission spectra of the planets by looking for excess absorption around the helium feature during the transits. We calculated a master out-of-transit spectrum by taking the weighted mean of the continuum normalized spectra with the data labeled according to the ephemerides given in Table~\ref{tab:transit_param}. For the targets of this study (i.e., the ones we observed) we assumed data during ingress and egress were in transit, but for WASP-107b we used the exact frames that \citet{kirk20} identified in their analysis for direct comparison. Note that the handful of frames that were obtained during telescope or software glitches were excluded from our analysis.

Individual transmission spectra for each planet were calculated as $T = 1 - F_{in_i}/F_{out}$ for each in transit spectrum. In this convention for the transmission spectrum the continuum is around zero and excess planetary absorption gives positive numbers. These individual transmission spectra were then shifted into the planet's rest frame according to the planetary orbital motion as calculable from Tables~\ref{tab:modeling_parameters} and \ref{tab:transit_param} before co-adding to make a single planetary transmission spectrum. 

This data processing was performed independently for NIRSPEC orders 70 and 71. The two resulting transmission spectra for each data set were then resampled to a common wavelength grid and averaged to give a combined spectrum. Figure \ref{fig:WASP107_analysis} shows the final averaged transmission spectrum for WASP-107b compared to the results of \citet{kirk20}, which were derived for order 70 only. We achieve excellent agreement with their detection of a large signal, which gives confidence in our pipeline. 

Figure \ref{fig:spectra} shows the measured transmission spectra for our targets including example forward models for helium absorption in the planets' upper atmospheres (the models are described in detail in \S\ref{sec:model}). As mentioned in the Introduction, we do not detect the signature of helium absorption in any of our targets. The scatter in the data is generally consistent with the expectations from the photon-limited error bars except near the positions of telluric absorption features where correlated residuals can be seen. The variability of the telluric absorption lines introduces increased noise in the transmission spectrum because the averaged out-of-transit spectrum doesn't divide out of all the individual spectra identically. Nevertheless, telluric contamination is not a significant concern for our investigation because no telluric absorption lines line up with the expected position of the helium feature. Furthermore, our division of the telluric absorption spectrum is optimized over many lines across a wide spectral bandpass (i.e., the entirety of each order). We also remind the reader that telluric emission lines are precisely removed using the alternate nod position spectra.

Prior to averaging, the data in order 71 had lower signal-to-noise and larger residuals because \texttt{Molecfit} had more trouble fitting the tellurics in that order as compared to order 70. Some nights are also better than others in terms of telluric absorption. For example, the second night for GJ\,9827d yielded higher signal-to-noise data and cleaner residuals from the telluric correction due to less telluric variability.  Our weighted combination of the order 70 and 71 data accounted for the differences in signal-to-noise, and we used the second dataset for GJ\,9827d to constrain our atmospheric models (see \S\ref{sec:model}).

\begin{figure}[t!]
\begin{center}
\includegraphics[width=\linewidth]{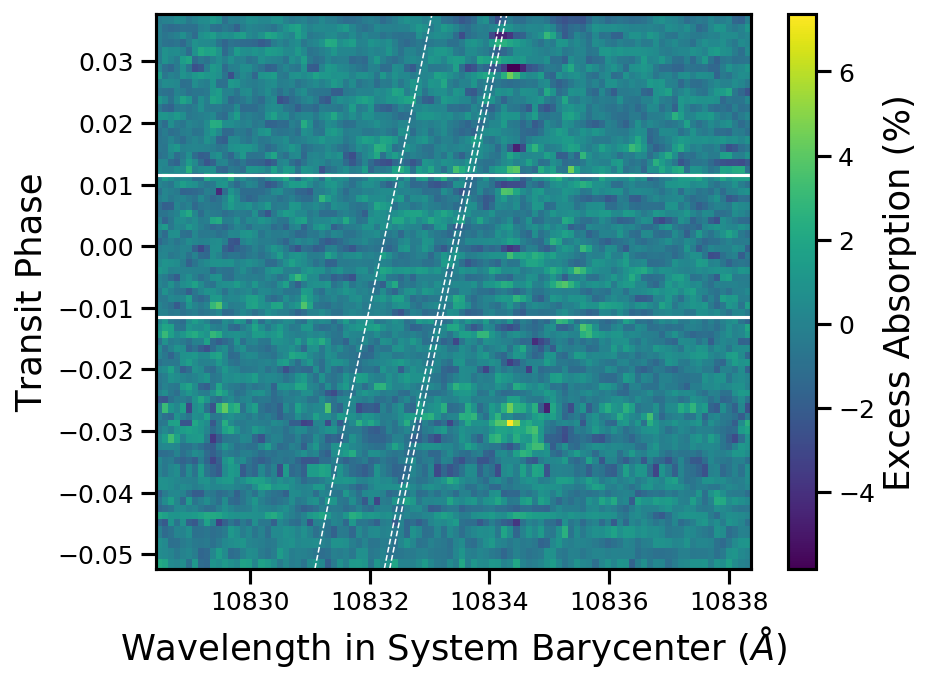}
\caption{\textbf{Excess absorption around the 10,833\,\AA\ feature for GJ\,1214b through time.} The excess absorption for order 70 is plotted in rest frame of the system barycenter as a function of phase. The horizontal lines indicate first and fourth contact of the transit. The dashed lines indicate the expected trace of the helium triplet lines due to the planet's orbital motion.} 
\label{fig:rv_search}
\end{center}
\end{figure}

As another test for helium absorption, we plotted the time series of data in the system barycenter frame relative to the master out-of-transit spectrum and searched for signals that were not perfectly centered on the transits. Spectral absorption from the upper atmospheres of transiting planets do not have to correspond to the white light transit of the bulk planet. For example, GJ\,436b shows a long tail of neutral hydrogen absorption at Lyman\,$\alpha$ \citep{ehrenreich15}, and \citet{kirk20} found that the helium absorption from WASP-107b began after ingress and continued after egress. Figure \ref{fig:rv_search} shows the time-series data for GJ\,1214b with the expected trace due to the planet's orbital motion. Neither GJ\,1214b nor any of the other planets in our sample showed evidence for a signal not centered on the transit.

\section{Atmosphere Modeling} \label{sec:model}
We used the obtained transmission spectra around the He 10,833\,\AA\ absorption signal to constrain the atmospheric properties of GJ\,1214b, HD\,97658b, and GJ\,9827d. Using methods similar to those described in \citet{oklopcic18} and \citet{mansfield18}, we modeled the upper atmosphere of each planet as a spherically symmetric hydrogen-helium envelope extending to altitudes of several planetary radii (up to the Roche radius of the system). The assumed atmospheric density profile was that of an isothermal Parker wind, in which the atmosphere is close to hydrostatic at low altitudes, but has a radial velocity component, increasing with altitude. An isothermal Parker wind model has two free parameters, the temperature of the upper atmosphere (thermosphere) and the density normalization, which can be linked to the atmospheric mass-loss rate and thus measures the total mass escaping from the planet per unit time.

\begin{figure}[t!]
\begin{center}
\includegraphics[width=\linewidth]{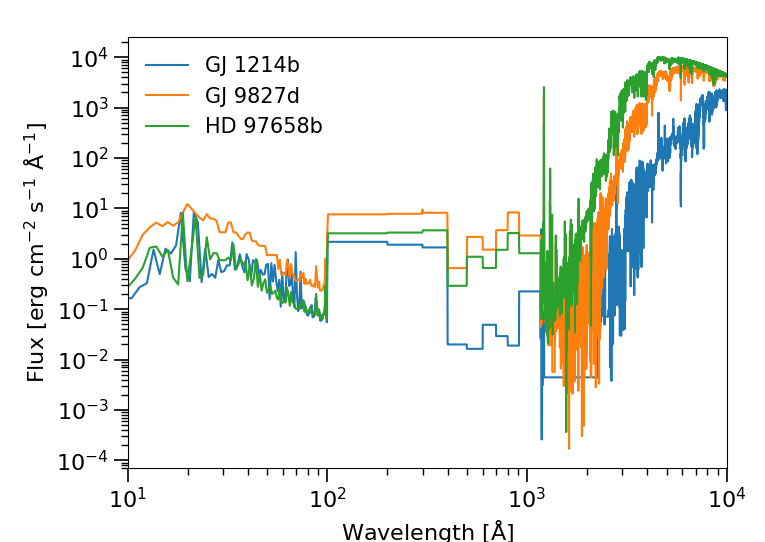}
\caption{\textbf{Stellar flux at the orbital distance of each planet used in our atmospheric modeling.} The SEDs of GJ\,1214 and HD\,97658 were obtained from the MUSCLES survey \citep{france16}. The SED of GJ\,9827 is modeled after a MUSCLES spectrum of a star of the same spectral type, with the EUV part reconstructed based on the Ly$\alpha$ observations by Carleo et al. (submitted).} 
\label{fig:sed}
\end{center}
\end{figure}

We perform radiative transfer calculations for each model atmosphere to obtain non-local thermodynamic equilibrium populations (as functions of altitude) of neutral and ionized hydrogen, and helium atoms in the ground, the metastable (2$^3$S), and the (singly) ionized state. For GJ\,1214 and HD\,97658, we use the spectral energy distributions (SEDs) of these stars obtained from the MUSCLES survey \citep{france16, youngblood16, loyd16}. For GJ\,9827, we reconstruct the EUV part of the spectrum using the observed Ly\,$\alpha$ flux (Carleo et al., submitted) and the scaling relations from \citet{Linsky2014}; the spectrum at longer wavelengths is modeled after the MUSCLES spectrum of HD\,85512, a star of the same spectral type as GJ\,9827. The estimated EUV fluxes incident on the planets are shown in Figure~\ref{fig:sed}. More details on the radiative transfer methods, including various reaction rate coefficients used in our calculations, can be found in \citet{oklopcic18}. 

Using the obtained 1D density profile of metastable helium and assuming a spherically symmetric planetary atmosphere, we computed a mid-transit transmission spectrum at wavelengths around 10,833\,\AA\ for a grid of model atmospheres for each planet. Each grid spans a broad range of values in thermospheric temperature (2,000 -- 8,000 K) and mass-loss rates ($10^6$ -- $10^{11}$ g s$^{-1}$). Our nominal model grid assumed roughly solar composition (10\% helium number fraction) because this is the expected composition of the wind (see \S\ref{sec:interior} and \ref{sec:discussion}). We also calculated model grids for GJ\,1214b with the helium fraction by number ranging from 5\% to 20\%, i.e.\ from subsolar to supersolar values, to explore how changing the helium abundance in the wind would impact our results.

By treating the properties of the Parker wind as free parameters, we acknowledge the current lack of understanding of how the properties of planetary winds depend on a whole range of planetary and stellar properties. Thus, we avoid making (potentially unfounded) assumptions about the wind properties. Alternatively, we could have assumed that the mass-loss rate is related to the planet's size and stellar EUV flux through the energy-limited formula. However, that would require us to assume a heating efficiency, which is unknown. Furthermore, the energy-limited approximation ignores a range of potentially important aspects of the problem, such as the existence of a planetary magnetic field or the interaction with the stellar wind, 
which are also unconstrained. By keeping the wind parameters free, we are being as agnostic as possible about all these issues. Basically, the main assumption that we make is related to the range of the parameter space that we explore. But, we compensate for that by exploring a wide range of mass-loss rates and temperatures around the values predicted by the results from the literature \citep[][and see \S\ref{sec:discussion}]{salz16}.



\begin{figure*}[t!]
\begin{center}
\includegraphics[width=\linewidth]{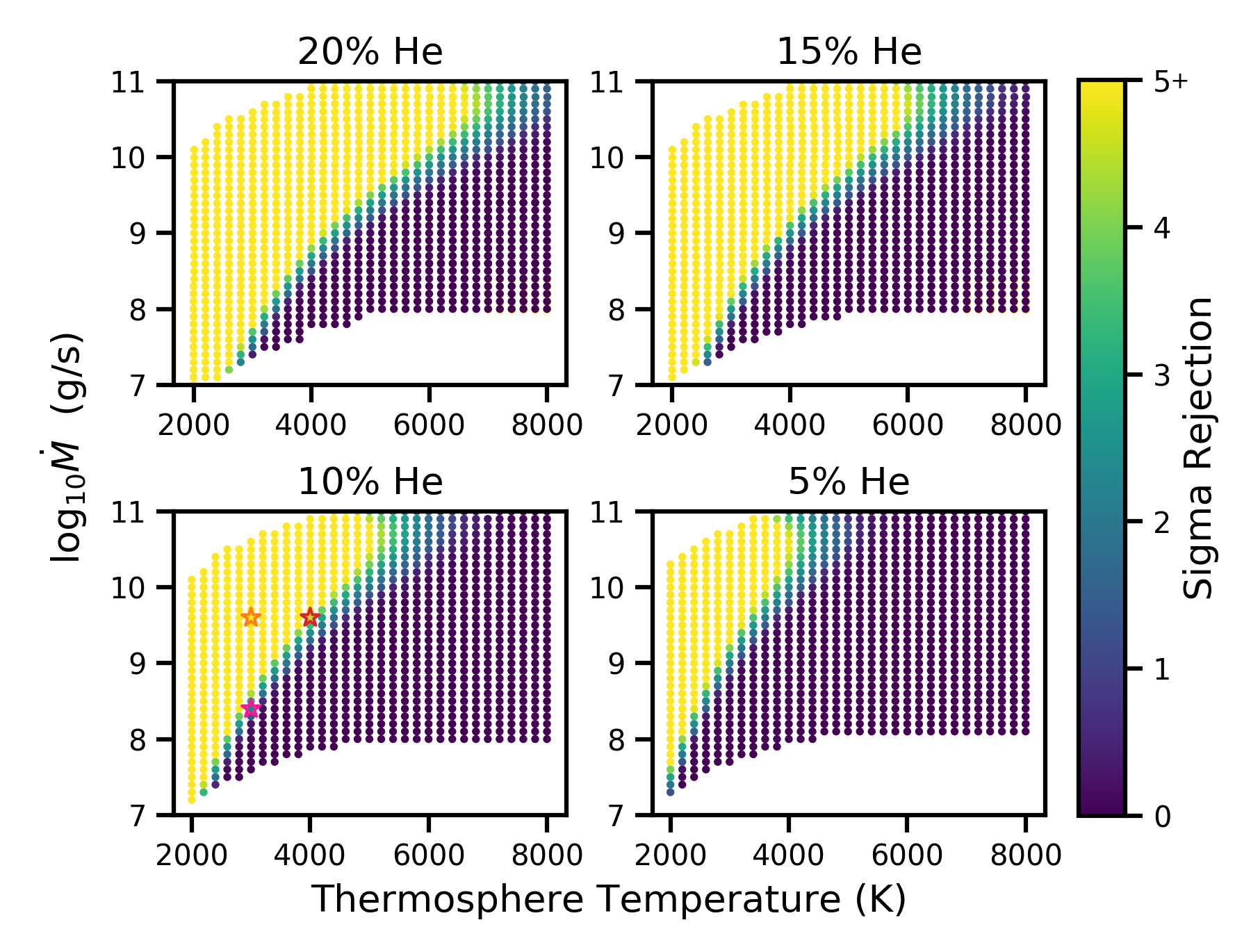}
\caption{\textbf{Map showing the statistical significance of the deviation from a good fit for the comparison of the models to our observations of GJ\,1214b, as a function of the thermospheric temperature and mass loss rate.} The different panels represent different abundances of helium in the wind as labeled at the top of each plot (values are given in percent by number). Regions of parameter space that fall outside the bounds of the \citet{salz16} simulations for a range of close-in planets are not plotted. The orange, red, and magenta stars represent the parameters of models shown in Figure~\ref{fig:spectra}, each filled with their respective significance. The orange star corresponds to the nominal planet-specific predictions of \citet{salz16}.} 
\label{fig:GJ1214_models}
\end{center}
\end{figure*}

We compared each simulated spectrum to the data to determine which parts of model parameter space could be excluded by our non-detection of the helium feature. To do this we first transformed the transmission models into their equivalent excess absorption. Then we broadened the models to the R\,=\,25,000 instrumental response and resampled them to the wavelengths of the combined (order 70 and 71) observed data set. We then identified the optimal window around the 10,833\,\AA\ feature for each model by maximizing the cumulative distribution function of the model absorption (defined at each pixel point) over the standard deviation of the observed data (defined in a constant $\pm1.5$\,\AA\ region around the feature). The window was centered on the middle of the two stronger lines of the triplet feature to maximize the contrast. The observed data region to use for testing the models was thus chosen as a compromise between sampling the true spread in the data assuming the null hypothesis and retaining information in the relevant region assuming the contrary. Following the optimization of the test window, the corresponding (maximal) rejection of the model given the data was found by its relation to the cumulative distribution function.

\begin{figure}[t!]
\begin{center}
\includegraphics[width=\linewidth]{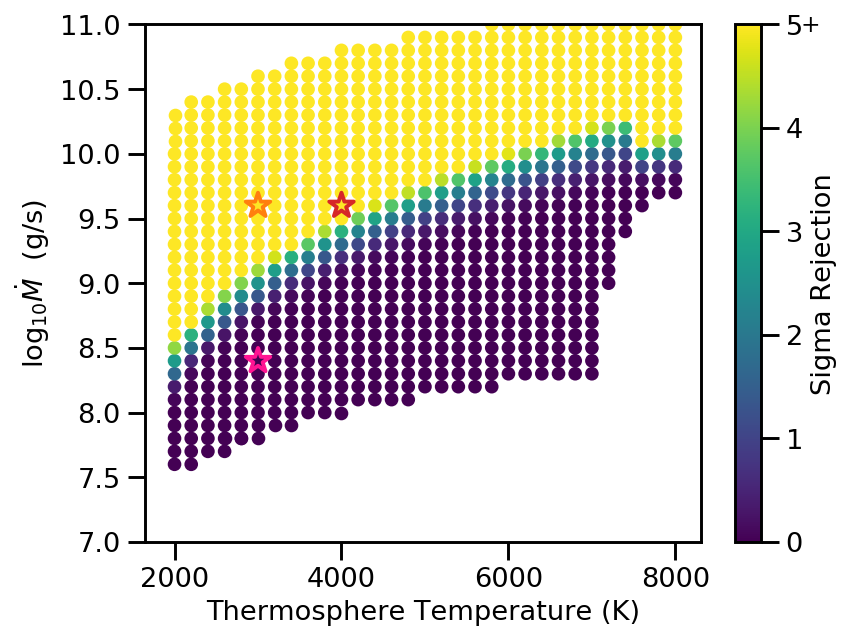}
\caption{\textbf{Similar to Figure~\ref{fig:GJ1214_models}, except for GJ\,9827d and limited to solar composition models.}} 
\label{fig:GJ9827_1126_model_rejection}
\end{center}
\end{figure}

\begin{figure}[t!]
\begin{center}
\includegraphics[width=\linewidth]{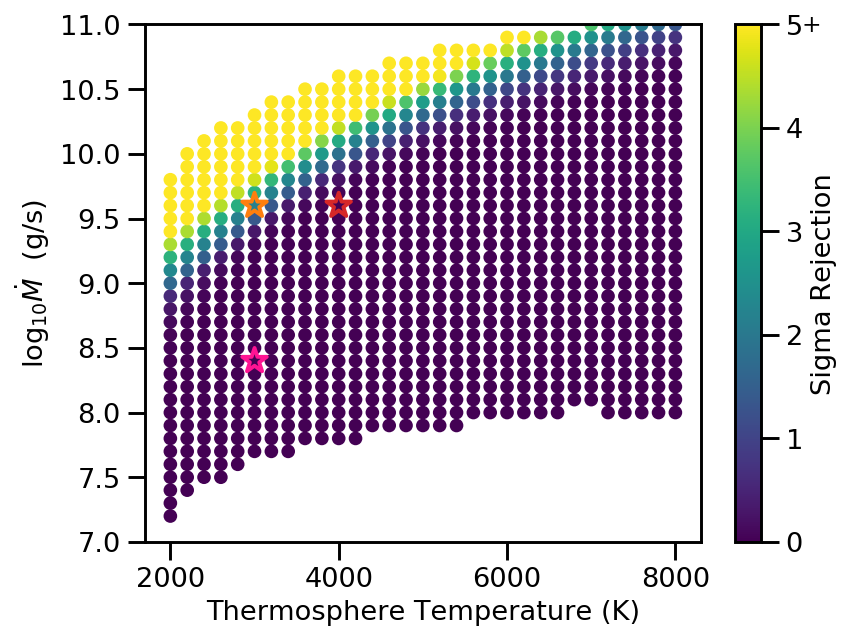}
\caption{\textbf{Similar to Figure~\ref{fig:GJ1214_models}, except for HD\,97658b and limited to solar composition models.}} 
\label{fig:HD97658_model_rejection}
\end{center}
\end{figure}

Figure \ref{fig:GJ1214_models} shows the map of the statistical deviation from our data for the GJ\,1214b model grids. As expected, the strength of the triplet feature is strongly correlated with the helium abundance in the wind. Therefore our non-detection rules out a larger part of the parameter space for high helium abundances, but is correspondingly less constraining for lower helium abundances. Figures \ref{fig:GJ9827_1126_model_rejection} and \ref{fig:HD97658_model_rejection} show the same maps for solar composition model grids for GJ\,9827d and HD\,97658b, respectively. We compared our model grid to the second observation of GJ\,9827d because it was the higher quality dataset of the two we obtained.

Notably, the contours for GJ\,1214b in Figure \ref{fig:GJ1214_models} reverse direction at high mass loss rates. This is because at a fixed temperature and at relatively low mass loss rate, increasing the rate increases the overall density, and consequently, the density of the metastable helium, which causes stronger absorption. However, after reaching a certain mass-loss rate value, the atmospheric density becomes so high that the EUV photons (which ionize helium and thus populate the metastable state) cannot penetrate all the way to the bottom of the atmosphere, so the lowest atmospheric layers become more and more depleted of metastable helium as the product of the mass-loss rate, and thus density, increases.

\begin{figure*}[t!]
\begin{center}
\includegraphics[width=\linewidth]{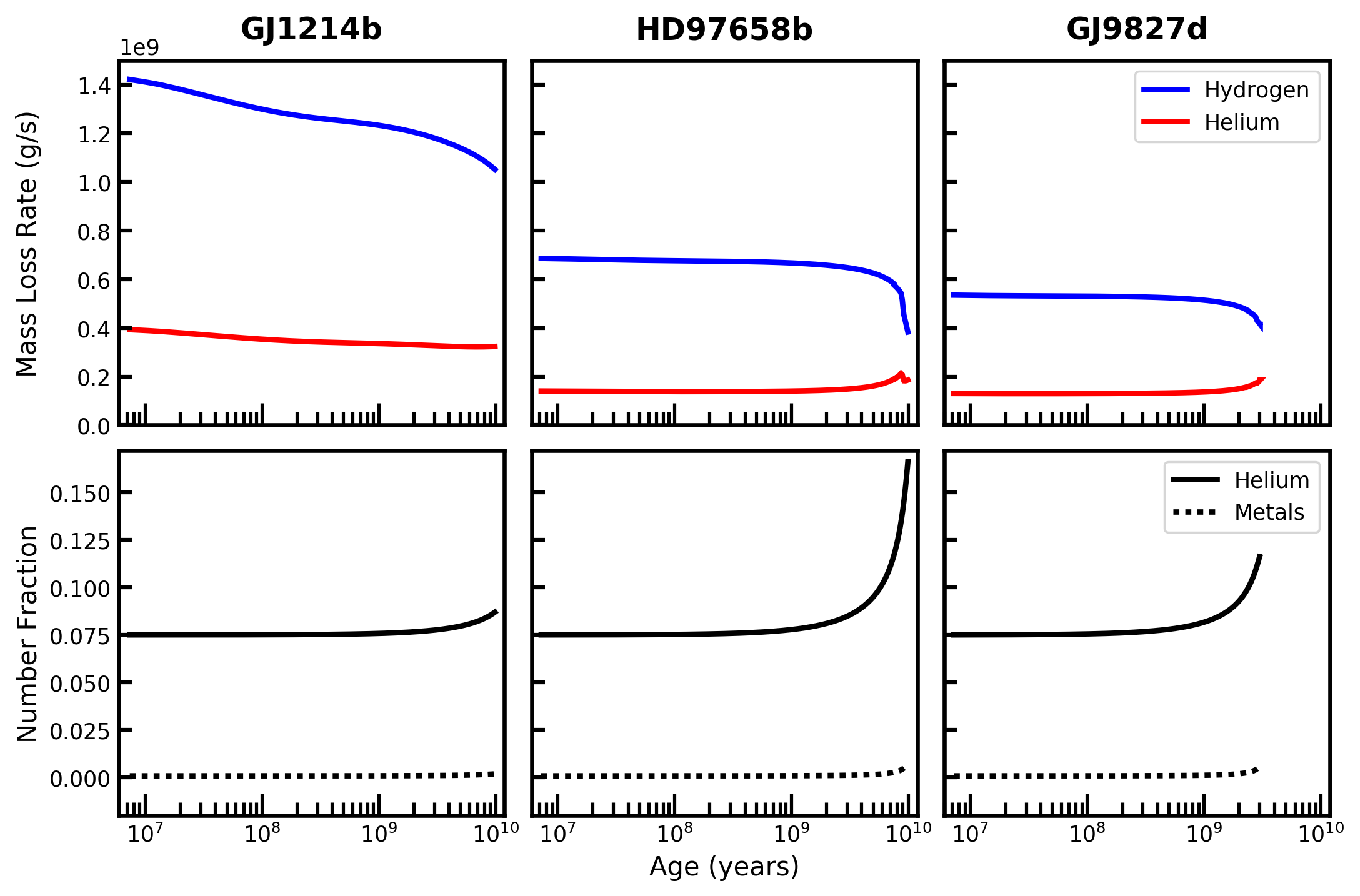}
\caption{\textbf{Hydrogen and helium mass loss rates (top row), as well as bulk envelope compositions (bottom row) for GJ\,1214b, HD\,97658b, and GJ\,9827d.} These models have initial envelope fractions of 0.022, 0.0063, and 0.003, respectively. At 10\,Gyr, the models for GJ\,1214b and HD\,97658b had radii consistent with their currently measured values (see Table~\ref{tab:modeling_parameters}). We were unable to simulate GJ\,9827d for 10\,Gyr due to extreme envelope depletion. However, the model evolved to match the present-day radius of this planet by 3\,Gyr. All models had homopause temperatures of 3,000\,K and heating efficiency factors of $\eta$ = 0.10.} 
\label{fig:InteriorStruct}
\end{center}
\end{figure*}


\section{Interior Structure Modeling} \label{sec:interior}
Fractionation between different chemical species could lead to atmospheric loss with a wind composition that is different from the bulk of the envelope. The planets in our sample are expected to be in the hydrodynamic mass loss regime where the large-scale loss of hydrogen drags heavier species along \citep[][and see \S\ref{sec:discussion}]{salz16}. Nevertheless, diffusive separation that takes place between the homopause (where the atmosphere is no longer expected to be well-mixed due to eddy diffusion) and the sonic point (where the flow composition becomes fixed) can reduce the amount of heavier species that are present in the wind. Helium is at risk of fractionation because it is four times heavier than the atomic hydrogen that makes up the bulk of the wind. For example, the energy-limited escape of hydrogen from modern Earth is incapable of pulling along helium and other heavy species \citep{catling17}. Mass fractionation in a hydrodynamic wind is enhanced for lower temperatures and lower mass-loss rates.

\citet{hu15} suggested that mass-dependent fractionation could lead to preferential loss of hydrogen and a corresponding enhancement in helium abundance in the bulk atmospheres of Neptune-size planets. If our targets are in a regime where the relative abundance of helium in their bulk was increasing, it would necessarily mean that the winds are actually depleted in helium, thus making our observation more difficult.

We performed coupled interior structure and atmospheric mass loss calculations to explore this phenomenon and provide further context for the non-detection of helium in the upper atmospheres of our targets. In this modeling, we consider a scenario where the planets consist of an Earth composition rocky core surrounded by a primordial hydrogen-helium envelope. By matching interior structure models to the observed characteristics of GJ\,1214b, HD\,97658b, and GJ\,9827d, we can constrain the planetary envelope fractions and predict the hydrogen and helium mass loss rates as a boundary condition for the atmosphere models. Furthermore, we can show how the bulk envelope fractions and wind compositions of these planets evolve over the course of billions of years.

We performed these calculations using the Modules for Experiments in Stellar Astrophysics (\texttt{MESA}) code \citep[v12778;][]{Paxton2011, Paxton2013, Paxton2015, Paxton2018, Paxton2019} and the coupled interior structure and atmospheric mass loss models outlined in \cite{Malsky_2020}. We instituted the regime of hydrogen and helium escape from \cite{hu15} to model the coupled thermal, mass-loss, and compositional evolution of hydrogen-helium envelopes surrounding our sub-Neptune size targets. More details of our calculations can be found in \citet{Malsky_2020}.

When the incident EUV radiation on a planet is extremely high the escaping wind is transonic and increases in incident flux do not drive further escape. Instead, the deposited energy is converted into translational and thermal energy in the atmosphere. To account for this we included the reduction factor from \cite{Johnson_2013}. This reduction factor reduces the mass loss rate when the incident EUV heating rate is above the critical heating rate of the planet. This results in the relatively constant mass loss rates shown in Figure~\ref{fig:InteriorStruct}. From early evolution to their present ages, the reduction factors of GJ\,1214b, HD\,97658b, and GJ\,9827d range from 1000 to 50, 20 to no reduction (1), and 100 to 20, respectively. The early evolution reduction factors are quite large in the case of these planets due to their small orbital separations and the exponential decrease of EUV flux as their host stars evolve \citep{2011A&A...532A...6S}.

We parameterized our grid of models as follows. The planetary masses, orbital separations, and host star characteristics were taken from Table~\ref{tab:modeling_parameters}. For each planet, we ran a grid with varying values for the EUV heating efficiency factor ($\eta$), initial envelope mass fraction, homopause temperature, and masses. All models started with an initial solar composition ($X=0.74$, $Y=0.24$, $Z=0.02$), and evolved from a `hot-start' to 10\,Gyr. After 6.0\,Myr of evolution, hydrodynamic mass loss was turned on. All other model parameters are identical to the full model description in \cite{Malsky_2020}.

We define the planetary radius to be at 1.0\,mbar, roughly the pressure level corresponding to the observed transit radii \citep{2009ApJ...702.1413M}. \texttt{MESA} does not calculate atmospheric structure down to these low pressures. Therefore, we extrapolate from the outermost zone in our models, assuming an isothermal temperature profile and constant values for gravity and the mean molecular mass.

The planet envelope has constant elemental abundances throughout each zone modeled in \texttt{MESA}. Mass is lost via a wind at the homopause radius. We defined the homopause radius as the location where the H-He binary diffusion coefficient is equal to the eddy diffusion coefficient. In general, the homopause radius is approximately 25\% to 60\% larger than the transit radius, with larger differences for smaller-mass planets with lower surface gravity. Our assumption that the wind is launched at the homopause is somewhat inconsistent with the predictions of the location of the thermosphere for our targets by \citet{salz16}, who suggest that the $\tau = 1$ level for high-energy photons is at radii $<$\,1.1\,R$_{p}$. However, this shouldn't have a strong impact on our results because the sonic points for the planets all occur at much larger radii ($>$\,3\,R$_{p}$). We adopt a nominal homopause temperature of 3,000\,K for all the planets to be consistent with the predictions of \citet{salz16}.



Figure~\ref{fig:InteriorStruct} shows the mass loss rates and bulk envelope evolution of three models, with radii matched for our individual targets. We find that the bulk atmospheric composition of GJ\,1214b does not change significantly from its initial solar abundance when evolved with hydrodynamic mass loss. In order to achieve the observed radius, the GJ\,1214b model requires an envelope approximately 0.81-0.96\% of the total planet mass \citep[depending on the system's uncertain age of 3 -- 10\,Gyr,][]{charbonneau09}, which precludes the bulk composition of the envelope from becoming significantly enhanced in helium. This result is robust against variations in age, homopause temperature, and planet mass.

HD\,97658b and GJ\,9827d have small enough transit radii that the bulk compositions of their envelopes may change through hydrodynamic mass loss. In particular, HD\,97658b may have a bulk envelope composition that is modestly enhanced in helium if it has been subject to fractionated mass loss over billions of years. For the model shown in Figure~\ref{fig:InteriorStruct}, 56\% of the envelope mass was lost, and the envelope helium number fraction increased from 0.075 to 0.168 over 10\,Gyr. However, \citet{henry11} estimated the age of the HD\,97658 system to be $\lesssim$7\,Gyr using three different indicators, including their measurement of 38.5\,days for the rotation period of the host star. \citet{guo20} found a shorter rotation period of 34\,days and also an activity cycle of 9.6\,yr, which both point to even younger ages. Therefore, the two times enhancement (depletion) of helium for the bulk (wind) of this planet is likely an overestimate because the fractionation is expected to ramp up toward later ages as the stellar high-energy flux and overall loss rate decrease.


Although GJ\,9827d has likely lost a significant fraction of its envelope, we predict that the bulk envelope composition is at most mildly enhanced in helium. For the model shown in Figure~\ref{fig:InteriorStruct}, GJ\,9827d lost over 84\% of its total envelope mass, and increased in helium number fraction over 3 Gyr to 0.117, corresponding to helium mass fractions going from 0.240 to 0.303. This is due to the fact that the planet is so highly irradiated that hydrogen and helium are lost in approximately equal proportion to their abundances in the bulk of the planetary envelope.

For models with similar radii to  GJ\,1214b, HD\,97658b, and GJ\,9827d, and a mass loss efficiency factor of $\eta$=0.10, we find average hydrogen mass loss rates of approximately 1.1 $\times$ 10$^9$, 3.8 $\times$ 10$^8$, and  4.2$\times$ 10$^8~\mathrm{g\,s^{-1}}$, respectively. Helium mass loss rates for the same set are 3.2 $\times$ 10$^8$, 1.8 $\times$ 10$^8$ and 1.8 $\times$ 10$^8~\mathrm{g\,s^{-1}}$, respectively.

Numerical instabilities prevented some of the models with the lowest initial envelope mass fractions from evolving for the full 10\,Gyr. In particular, the mass and radius of GJ\,9827d are at the boundaries of the equation of state (EOS) tables in \texttt{MESA}, and are therefore at the limit of what we can simulate. A more detailed discussion of \texttt{MESA}'s capability of modeling highly irradiated sub-Neptunes can be found in  \cite{Malsky_2020}.

\section{Discussion} \label{sec:discussion}
The key question for this paper is, what can the non-detections of the helium feature tell us about the atmospheric properties of sub-Neptune size planets? Our coupled interior structure and atmospheric mass loss calculations indicate that the compositions of the bulk atmospheres of these planets shouldn't be significantly altered from their original state. Therefore, our targets will have helium abundances roughly corresponding to the solar value in their bulk and winds if sub-Neptune size planets accrete primary atmospheres during formation. While helium-enhanced winds would be more detectable, our observations for GJ\,1214b and GJ\,9827d were sensitive to upper atmospheres with solar composition for a wide range of temperatures and mass loss rates. Thus the question becomes, what temperatures and mass loss rates were expected for these planets?

\citet{salz16} presented planet-specific simulations of atmospheric escape for GJ\,1214b and HD\,97658b using a coupled 1D radiative-hydrodynamical simulation code. They predict thermosphere temperatures of approximately 2,500 and 3,500\,K and mass loss rates of $5 \times 10^9$ and $3 \times 10^9$\,g\,s$^{-1}$ for these two planets, respectively. Since these predictions are so similar, and because all three planets have comparable incident EUV fluxes (see Figure~\ref{fig:sed}) and physical parameters (see Figure~\ref{fig:sample}), we take average values of T\,=\,3,000\,K and $\dot{M} = 4 \times 10^{9}$\,g\,s$^{-1}$ as nominal predictions for all the planets for the sake of simplicity. The models indicated by the solid orange lines and dots in Figures~\ref{fig:spectra}, \ref{fig:GJ1214_models} -- \ref{fig:HD97658_model_rejection}, and \ref{fig:delta_EUV} correspond to this nominal set of parameters. The red and magenta models correspond to changing the temperature or mass loss rate by an illustrative amount from this point in parameter space.

Note that the mass loss rates predicted by \citet{salz16} are higher than those predicted by our own calculations that were presented in \S\ref{sec:interior} by about a factor of four. This is probably due to our adoption of a low efficiency factor, while \citet{salz16} self consistently account for the physics that is captured in this term. Higher mass loss rates would reduce the fractionation between hydrogen and helium, thus supporting our assumption that the winds of our targets should be roughly solar composition if they originally began with primary atmospheres.

\begin{deluxetable*}{lccccc}
\tablecolumns{6}
\tablecaption{\label{tab:measurements}Atmospheric constraints for our targets}
\tablehead{
 \colhead{Planet} & \colhead{$\Delta$D (\%)}\tablenotemark{a} & \colhead{$\delta_{Rp}$ (km)}\tablenotemark{a} & \colhead{$H_{eq}$ (km)} &  \colhead{$\delta_{Rp}/H_{eq}$}\tablenotemark{a} & \colhead{Flux$_{\mathrm{XEUV}}$ (W m$^{-2}$)}\tablenotemark{b}}
\startdata
GJ\,1214b & 0.13 & 817 & 250 & 12 & 0.64\\
GJ\,9827d & 0.067 & 3,917 & 230 & 26 & 2.45\\
HD\,97658b & 0.21 & 13,426 & 181 & 84 & 1.11\\
\enddata
\tablenotetext{a}{Measurements are 90\% confidence (1.645$\sigma$) upper limits.}
\tablenotetext{b}{Determined by integrating our stellar models from 5 to 504\,\AA. We conservatively estimate that these values are accurate to within a factor of three.}
\end{deluxetable*}

As can be seen in the figures, the nominal predictions of \citet{salz16} and the surrounding parameter space can be ruled out for GJ\,1214b and GJ\,9827d at high confidence. The results for HD\,97658b are less constraining due to a combination of an unfavorable planet-to-star radius ratio, poor observing conditions limiting the signal-to-noise of the data, and the telluric contamination that is endemic to ground-based observations. Further observations of this target would be useful.

\begin{figure}[t!]
\begin{center}
\includegraphics[width=\linewidth]{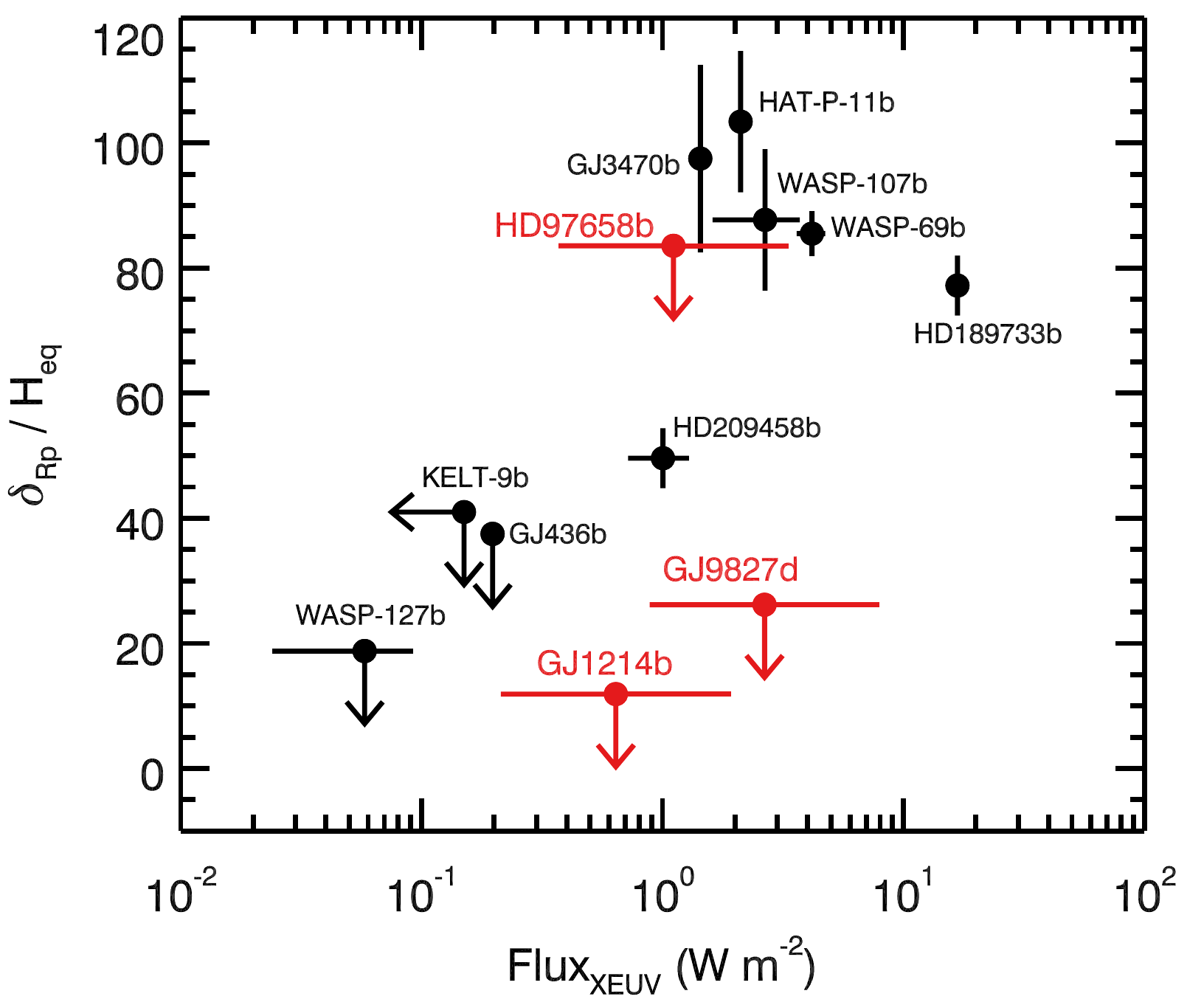}
\caption{\textbf{Equivalent height of the absorbing atmosphere as a function of high-energy instellation ($\lambda$ = 5 -- 504\,\AA) for planets with helium feature detections and upper limits.} The equivalent height of the absorbing atmosphere ($\delta_{Rp}$) is normalized by the scale height of the bulk atmosphere ($H_{eq}$). This metric was originally propsed by \citet{nortmann18}. The black points with errors are measurements taken from the literature. The GJ\,3470b data are from \citet{palle20}, and the rest are from the compilation of \citet{dossantos20}. The red points are measurements for our sub-Neptune targets. The arrows represent 90\% confidence upper limits.}
\label{fig:strength}
\end{center}
\end{figure}

We place our measurements in the context of other planets with published helium detections and upper limits in Figure~\ref{fig:strength}. Following \citet{nortmann18}, this figure plots the ``equivalent height'' of the absorbing atmosphere normalized by the scale height of the bulk atmosphere as a function of the high-energy instellation of the planets (``Flux$_{\mathrm{XEUV}}$'', $\lambda$ = 5 -- 504\,\AA). The measurements for our planets are tabulated in Table~\ref{tab:measurements}. The equivalent height of the absorbing atmosphere measurements we quote are 90\% confidence upper limits. This confidence limit was estimated by taking the median maximum transit depth of models for each planet that were inconsistent with the data at the approximately 1.65$\sigma$ level. We used the maximum transit depth of the models at their native (full) resolution because our relatively low-resolution observations ($R$\,=\,25k when $R$\,$>$\,50k is more typical for these observations) will make a given transit depth appear shallower, thus reducing the apparent height of the atmosphere from the true value. The results in Figure~\ref{fig:strength} for GJ\,1214b and GJ\,9827d do not follow the correlation that is seen for larger planets between the high-energy instellation and the normalized height of the helium absorption. This suggests that the upper atmospheres of our targets are fundamentally different than those of giant planets.


The lack of a detection and the limits we can set for three similar planets that orbit stars with a range of high-energy fluxes points to a problem with at least one fundamental assumption in our chain of hypotheses about the upper atmospheres of sub-Neptune planets. One possible explanation for the mismatch between the model predictions and the data is that the models are inaccurate. We previously showed that the combination of the \citet{salz16} and \citet{oklopcic18} models perfectly reproduced observations of helium absorption in the Neptune-sized exoplanet HAT-P-11b \citep{mansfield18}. However, the predictions of the temperature and escape rate in the \citet{salz16} model, and the population of the metastable level that gives rise to the helium feature in the \citet{oklopcic19} model both depend on the input stellar spectra in the EUV range. The EUV flux is usually reconstructed from the observed X-ray or UV fluxes, which are not precisely established for our targets or HAT-P-11b \citep{loyd16}. \citet[][see her Figure 7]{oklopcic19} has shown that reasonable variations in how the stellar EUV spectrum is reconstructed can lead to factor of three differences in the strength of the helium feature.

Figure~\ref{fig:delta_EUV} shows the result of changing the input stellar spectra in our radiative transfer calculations by almost an order of magnitude, which is the typical level of uncertainty in the reconstructed EUV spectral energy distributions \citep{Youngblood2019, Drake2020}. \citet{salz16} estimate that the uncertainties in the stellar fluxes generally lead to factor of three uncertainties in the mass-loss rates. Therefore, we also changed the mass loss rate by a factor of three along with the stellar EUV flux in these test calculations. Even in the low EUV flux case, which produces the weakest He absorption signal, the predicted signal is ruled out by the observations at 4$\sigma$ confidence. Nevertheless, this test shows that good characterization of the high-energy fluxes of stellar host stars are essential for understanding atmospheric loss in the sub-Neptune regime.

\begin{figure}[t!]
\begin{center}
\includegraphics[width=\linewidth]{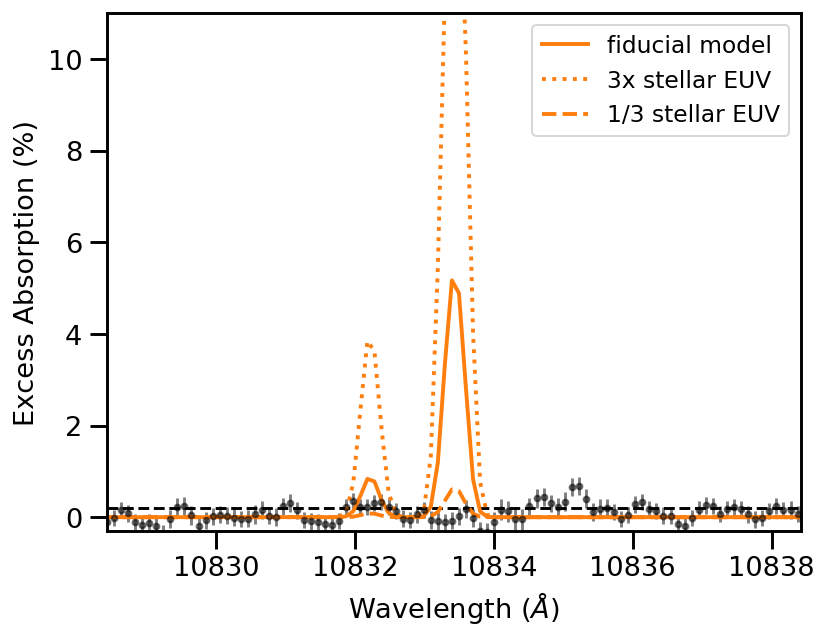}
\caption{\textbf{The transmission spectrum of GJ\,1214b compared to models using different stellar EUV flux levels.} The data for GJ\,1214b are plotted as grey points with errors. The black dashed line indicates the standard deviation of the data around the He feature. The solid orange model is created utilizing the fiducial model (T\,$=$\,3,000\,K, log\,$\dot{M}$ = 9.6) and the nominal stellar EUV flux. The dashed line model was created utilizing a factor of three less EUV flux and a lower mass loss rate of log\,$\dot{M}$ = 9.1. The dotted model was created utilizing a factor of three more EUV flux and a higher mass loss rate of log\,$\dot{M}$ = 10.1. The fiducial and high stellar EUV flux models are ruled out at very high confidence, while the lower stellar EUV flux model is ruled out at 4$\sigma$ confidence.}
\label{fig:delta_EUV}
\end{center}
\end{figure}

Another reason why our models might be inaccurate is that the lower temperatures of our sub-Neptune planets could impact the composition of their thermospheres. Both the \citet{salz16} and \citet{oklopcic18} models assume that all the hydrogen is in atomic form. This is likely a safe assumption for HAT-P-11b, where the thermosphere temperature was predicted to be $\sim$7,000\,K at the radial distance from the planet probed by the helium feature. In contrast, the thermosphere temperatures of our targets ($\sim$3,000\,K) are around the point that a significant amount of hydrogen could be in its molecular form. The presence of molecules will reduce the escape rate because heavier chemical species will have lower thermal velocities for a given temperature.

\citet{salz16} describe a test calculation for GJ\,1214b where they include molecule formation. They quote H$_{2}$ abundances of 10 -- 25\% throughout the thermosphere of the planet, and corresponding reductions in the mass loss rate of 15\%. This level of uncertainty in the mass loss rate is not significant for our work. Also, it is not clear if their calculations include the impact of molecular diffusion, which will cause the heaver H$_{2}$ to preferentially settle out, thus reducing its impact on the atmospheric escape. Nevertheless, more sophisticated models are needed to understand atmospheric mass loss in the low temperature and low surface gravity regime of sub-Neptune size planets.


Finally, the possibility that these planets have metal-rich atmospheres cannot be ruled out. Such atmospheres would have lower helium abundances, which reduces the strength of the helium feature (see Figure~\ref{fig:GJ1214_models}). Also, having more metals in the thermosphere could increase the cooling efficiency, thus leading to lower temperatures that will reduce the mass loss rate. We are not aware of any papers that study the details of atmospheric mass loss specifically for thick, metal-rich atmospheres. Hopefully our observational results motivate work in this area.

\section{Summary} \label{sec:summary}
To summarize, we set upper limits on the presence of the helium infrared triplet in high resolution transmission spectra of the sub-Neptune size planets GJ\,1214b, HD\,97658b, and GJ\,9827d. The nominal predictions of models for GJ\,1214b and GJ\,9827d are inconsistent with our non-detections at high significance if we assume these planets were born with primary atmospheres. These findings are somewhat robust (more so for GJ\,1214b than GJ\,9827d) to uncertainties in the expected temperatures, mass loss rates, and helium abundances of the planets' winds, and the EUV fluxes of their host stars. The results for these two planets are also inconsistent with the emerging trend of the detection of large helium features in giant planets with modest to to large high-energy instellation. The data quality for HD\,97658b was insufficient to detect the nominal model predictions due to the observations being obtained in poor conditions. More sophisticated models for the atmospheric evolution of cool, metal-rich atmospheres are needed to interpret these results. More precise characterization of the stellar EUV fluxes would also be valuable.

\acknowledgments
We thank the anonymous referee for comments that helped improve this paper. We thank James Kirk, Jon Otegi, B.J.\ Fulton, and Leonardo dos Santos for sharing data from their recent papers. We thank Greg Doppmann for his assistance with the observations, and we thank Elisabeth Newton for discussions about the radial velocity of GJ\,1214. We thank Allison Youngblood for helpful conversations regarding the stellar EUV reconstruction and Ilaria Carleo for sharing the results on the observed Ly\,$\alpha$ flux of GJ\,9827.

This work was supported by a NASA Keck PI Data Award, administered by the NASA Exoplanet Science Institute. Data presented herein were obtained at the W.\ M.\ Keck Observatory from telescope time allocated to the National Aeronautics and Space Administration through the agency's scientific partnership with the California Institute of Technology and the University of California. The Observatory was made possible by the generous financial support of the W.\ M.\ Keck Foundation.

A.O.\ acknowledges support by NASA through the NASA Hubble Fellowship grant HST-HF2-51443.001-A awarded by the Space Telescope Science Institute, which is operated by the Association of Universities for Research in Astronomy, Incorporated, under NASA contract NAS5-26555. E.M.-R.K.\ acknowledges support from the National Science Foundation under CAREER grant No.\ 1931736 and from the Research Corporation for Science Advancement through their Cottrell Scholar program. J.-M.D.\ acknowledges support from the Amsterdam Academic Alliance (AAA) Program, and the European Research Council (ERC) European Union's Horizon 2020 research and innovation program (grant agreement no.\ 679633; Exo-Atmos). L.A.R acknowledges support from the National Science Foundation under grant No. AST-1615315 and from the Research Corporation for Science Advancement through their Cottrell Scholar program. M.M.\ acknowledges funding from a NASA FINESST grant.

The authors wish to recognize and acknowledge the very significant cultural role and reverence that the summit of Mauna Kea has always had within the indigenous Hawaiian community. We are most fortunate to have the opportunity to conduct observations from this mountain.	

\facilities{Keck (NIRSPEC)}


\software{astropy \citep{Astropy2013}}







\begin{thebibliography}{}
\expandafter\ifx\csname natexlab\endcsname\relax\def\natexlab#1{#1}\fi

\bibitem[{{Adams} {et~al.}(2019){Adams}, {Gao}, {de Pater}, \&
  {Morley}}]{adams19}
{Adams}, D., {Gao}, P., {de Pater}, I., \& {Morley}, C.~V. 2019, \apj, 874, 61

\bibitem[{{Adams} {et~al.}(2008){Adams}, {Seager}, \&
  {Elkins-Tanton}}]{adams08}
{Adams}, E.~R., {Seager}, S., \& {Elkins-Tanton}, L. 2008, \apj, 673, 1160

\bibitem[{{Allart} {et~al.}(2018){Allart}, {Bourrier}, {Lovis}, {Ehrenreich},
  {Spake}, {Wyttenbach}, {Pino}, {Pepe}, {Sing}, \& {Lecavelier des
  Etangs}}]{allart18}
{Allart}, R., {Bourrier}, V., {Lovis}, C., {et~al.} 2018, Science, 362, 1384

\bibitem[{{Allart} {et~al.}(2019){Allart}, {Bourrier}, {Lovis}, {Ehrenreich},
  {Aceituno}, {Guijarro}, {Pepe}, {Sing}, {Spake}, \& {Wyttenbach}}]{allart19}
---. 2019, \aap, 623, A58

\bibitem[{{Alonso-Floriano} {et~al.}(2019){Alonso-Floriano}, {Snellen},
  {Czesla}, {Bauer}, {Salz}, {Lamp{\'o}n}, {Lara}, {Nagel},
  {L{\'o}pez-Puertas}, {Nortmann}, {S{\'a}nchez-L{\'o}pez}, {Sanz-Forcada},
  {Caballero}, {Reiners}, {Ribas}, {Quirrenbach}, {Amado}, {Aceituno},
  {Anglada-Escud{\'e}}, {B{\'e}jar}, {Brinkm{\"o}ller}, {Hatzes}, {Henning},
  {Kaminski}, {K{\"u}rster}, {Labarga}, {Montes}, {Pall{\'e}}, {Schmitt}, \&
  {Zapatero Osorio}}]{alonsofloriano19}
{Alonso-Floriano}, F.~J., {Snellen}, I.~A.~G., {Czesla}, S., {et~al.} 2019,
  \aap, 629, A110

\bibitem[{{Astropy Collaboration} {et~al.}(2013){Astropy Collaboration},
  {Robitaille}, {Tollerud}, {Greenfield}, {Droettboom}, {Bray}, {Aldcroft},
  {Davis}, {Ginsburg}, {Price-Whelan}, {Kerzendorf}, {Conley}, {Crighton},
  {Barbary}, {Muna}, {Ferguson}, {Grollier}, {Parikh}, {Nair}, {Unther},
  {Deil}, {Woillez}, {Conseil}, {Kramer}, {Turner}, {Singer}, {Fox}, {Weaver},
  {Zabalza}, {Edwards}, {Azalee Bostroem}, {Burke}, {Casey}, {Crawford},
  {Dencheva}, {Ely}, {Jenness}, {Labrie}, {Lim}, {Pierfederici}, {Pontzen},
  {Ptak}, {Refsdal}, {Servillat}, \& {Streicher}}]{Astropy2013}
{Astropy Collaboration}, {Robitaille}, T.~P., {Tollerud}, E.~J., {et~al.} 2013,
  \aap, 558, A33

\bibitem[{{Bean} {et~al.}(2010{\natexlab{a}}){Bean}, {Miller-Ricci Kempton}, \&
  {Homeier}}]{bean10b}
{Bean}, J.~L., {Miller-Ricci Kempton}, E., \& {Homeier}, D. 2010{\natexlab{a}},
  \nat, 468, 669

\bibitem[{{Bean} {et~al.}(2010{\natexlab{b}}){Bean}, {Seifahrt}, {Hartman},
  {Nilsson}, {Wiedemann}, {Reiners}, {Dreizler}, \& {Henry}}]{bean10a}
{Bean}, J.~L., {Seifahrt}, A., {Hartman}, H., {et~al.} 2010{\natexlab{b}},
  \apj, 713, 410

\bibitem[{{Bean} {et~al.}(2011){Bean}, {D{\'e}sert}, {Kabath}, {Stalder},
  {Seager}, {Miller-Ricci Kempton}, {Berta}, {Homeier}, {Walsh}, \&
  {Seifahrt}}]{bean11}
{Bean}, J.~L., {D{\'e}sert}, J.-M., {Kabath}, P., {et~al.} 2011, \apj, 743, 92

\bibitem[{{Benneke} {et~al.}(2019){Benneke}, {Wong}, {Piaulet}, {Knutson},
  {Lothringer}, {Morley}, {Crossfield}, {Gao}, {Greene}, {Dressing},
  {Dragomir}, {Howard}, {McCullough}, {Kempton}, {Fortney}, \&
  {Fraine}}]{benneke19}
{Benneke}, B., {Wong}, I., {Piaulet}, C., {et~al.} 2019, \apjl, 887, L14

\bibitem[{{Berta} {et~al.}(2012){Berta}, {Charbonneau}, {D{\'e}sert},
  {Miller-Ricci Kempton}, {McCullough}, {Burke}, {Fortney}, {Irwin}, {Nutzman},
  \& {Homeier}}]{berta12}
{Berta}, Z.~K., {Charbonneau}, D., {D{\'e}sert}, J.-M., {et~al.} 2012, \apj,
  747, 35

\bibitem[{{Bourrier} {et~al.}(2017){Bourrier}, {Ehrenreich}, {King},
  {Lecavelier des Etangs}, {Wheatley}, {Vidal-Madjar}, {Pepe}, \&
  {Udry}}]{bourrier17}
{Bourrier}, V., {Ehrenreich}, D., {King}, G., {et~al.} 2017, \aap, 597, A26

\bibitem[{{Catling} \& {Kasting}(2017)}]{catling17}
{Catling}, D.~C., \& {Kasting}, J.~F. 2017, {Atmospheric Evolution on Inhabited
  and Lifeless Worlds}

\bibitem[{{Chachan} \& {Stevenson}(2018)}]{chachan18}
{Chachan}, Y., \& {Stevenson}, D.~J. 2018, \apj, 854, 21

\bibitem[{{Charbonneau} {et~al.}(2009){Charbonneau}, {Berta}, {Irwin}, {Burke},
  {Nutzman}, {Buchhave}, {Lovis}, {Bonfils}, {Latham}, {Udry}, {Murray-Clay},
  {Holman}, {Falco}, {Winn}, {Queloz}, {Pepe}, {Mayor}, {Delfosse}, \&
  {Forveille}}]{charbonneau09}
{Charbonneau}, D., {Berta}, Z.~K., {Irwin}, J., {et~al.} 2009, \nat, 462, 891

\bibitem[{{Crossfield} {et~al.}(2019){Crossfield}, {Barman}, {Hansen}, \&
  {Frewen}}]{crossfield19}
{Crossfield}, I.~J.~M., {Barman}, T., {Hansen}, B., \& {Frewen}, S. 2019,
  Research Notes of the American Astronomical Society, 3, 24

\bibitem[{{dos Santos} {et~al.}(2020){dos Santos}, {Ehrenreich}, {Bourrier},
  {Allart}, {King}, {Lendl}, {Lovis}, {Margheim}, {Mel{\'e}ndez}, {Seidel}, \&
  {Sousa}}]{dossantos20}
{dos Santos}, L.~A., {Ehrenreich}, D., {Bourrier}, V., {et~al.} 2020, arXiv
  e-prints, arXiv:2007.06216

\bibitem[{{Dragomir} {et~al.}(2013){Dragomir}, {Matthews}, {Eastman},
  {Cameron}, {Howard}, {Guenther}, {Kuschnig}, {Moffat}, {Rowe}, {Rucinski},
  {Sasselov}, \& {Weiss}}]{dragomir13}
{Dragomir}, D., {Matthews}, J.~M., {Eastman}, J.~D., {et~al.} 2013, \apjl, 772,
  L2

\bibitem[{{Drake} {et~al.}(2020){Drake}, {Kashyap}, {Wargelin}, \&
  {Wolk}}]{Drake2020}
{Drake}, J.~J., {Kashyap}, V.~L., {Wargelin}, B.~J., \& {Wolk}, S.~J. 2020,
  \apj, 893, 137

\bibitem[{{Ehrenreich} {et~al.}(2015){Ehrenreich}, {Bourrier}, {Wheatley},
  {Lecavelier des Etangs}, {H{\'e}brard}, {Udry}, {Bonfils}, {Delfosse},
  {D{\'e}sert}, {Sing}, \& {Vidal-Madjar}}]{ehrenreich15}
{Ehrenreich}, D., {Bourrier}, V., {Wheatley}, P.~J., {et~al.} 2015, \nat, 522,
  459

\bibitem[{{Elkins-Tanton} \& {Seager}(2008)}]{elkins-tanton08}
{Elkins-Tanton}, L.~T., \& {Seager}, S. 2008, \apj, 685, 1237

\bibitem[{{France} {et~al.}(2016){France}, {Loyd}, {Youngblood}, {Brown},
  {Schneider}, {Hawley}, {Froning}, {Linsky}, {Roberge}, {Buccino},
  {Davenport}, {Fontenla}, {Kaltenegger}, {Kowalski}, {Mauas}, {Miguel},
  {Redfield}, {Rugheimer}, {Tian}, {Vieytes}, {Walkowicz}, \&
  {Weisenburger}}]{france16}
{France}, K., {Loyd}, R.~O.~P., {Youngblood}, A., {et~al.} 2016, \apj, 820, 89

\bibitem[{{Fulton} \& {Petigura}(2018)}]{fulton18}
{Fulton}, B.~J., \& {Petigura}, E.~A. 2018, \aj, 156, 264

\bibitem[{{Fulton} {et~al.}(2017){Fulton}, {Petigura}, {Howard}, {Isaacson},
  {Marcy}, {Cargile}, {Hebb}, {Weiss}, {Johnson}, {Morton}, {Sinukoff},
  {Crossfield}, \& {Hirsch}}]{fulton17}
{Fulton}, B.~J., {Petigura}, E.~A., {Howard}, A.~W., {et~al.} 2017, \aj, 154,
  109

\bibitem[{{Gaia Collaboration}(2018)}]{gaiadr2}
{Gaia Collaboration}. 2018, VizieR Online Data Catalog, I/345

\bibitem[{{Gaidos} {et~al.}(2020){Gaidos}, {Hirano}, {Mann}, {Owens}, {Berger},
  {France}, {Vanderburg}, {Harakawa}, {Hodapp}, {Ishizuka}, {Jacobson},
  {Konishi}, {Kotani}, {Kudo}, {Kurokawa}, {Kuzuhara}, {Nishikawa}, {Omiya},
  {Serizawa}, {Tamura}, \& {Ueda}}]{gaidos20}
{Gaidos}, E., {Hirano}, T., {Mann}, A.~W., {et~al.} 2020, \mnras, 495, 650

\bibitem[{{Gao} \& {Benneke}(2018)}]{gao18}
{Gao}, P., \& {Benneke}, B. 2018, \apj, 863, 165

\bibitem[{{Gillon} {et~al.}(2014){Gillon}, {Demory}, {Madhusudhan}, {Deming},
  {Seager}, {Zsom}, {Knutson}, {Lanotte}, {Bonfils}, {D{\'e}sert}, {Delrez},
  {Jehin}, {Fraine}, {Magain}, \& {Triaud}}]{gillon14}
{Gillon}, M., {Demory}, B.~O., {Madhusudhan}, N., {et~al.} 2014, \aap, 563, A21

\bibitem[{{Guo} {et~al.}(2020){Guo}, {Crossfield}, {Dragomir}, {Kosiarek},
  {Lothringer}, {Mikal-Evans}, {Rosenthal}, {Benneke}, {Knutson}, {Dalba},
  {Kempton}, {Henry}, {McCullough}, {Barman}, {Blunt}, {Chontos}, {Fortney},
  {Fulton}, {Hirsch}, {Howard}, {Isaacson}, {Matthews}, {Mocnik}, {Morley},
  {Petigura}, \& {Weiss}}]{guo20}
{Guo}, X., {Crossfield}, I. J.~M., {Dragomir}, D., {et~al.} 2020, \aj, 159, 239

\bibitem[{{Henry} {et~al.}(2011){Henry}, {Howard}, {Marcy}, {Fischer}, \&
  {Johnson}}]{henry11}
{Henry}, G.~W., {Howard}, A.~W., {Marcy}, G.~W., {Fischer}, D.~A., \&
  {Johnson}, J.~A. 2011, arXiv e-prints, arXiv:1109.2549

\bibitem[{{Horne}(1986)}]{horne86}
{Horne}, K. 1986, \pasp, 98, 609

\bibitem[{{Howard} {et~al.}(2011){Howard}, {Johnson}, {Marcy}, {Fischer},
  {Wright}, {Henry}, {Isaacson}, {Valenti}, {Anderson}, \&
  {Piskunov}}]{howard11}
{Howard}, A.~W., {Johnson}, J.~A., {Marcy}, G.~W., {et~al.} 2011, \apj, 730, 10

\bibitem[{{Hu} {et~al.}(2015){Hu}, {Seager}, \& {Yung}}]{hu15}
{Hu}, R., {Seager}, S., \& {Yung}, Y.~L. 2015, \apj, 807, 8

\bibitem[{Johnson {et~al.}(2013)Johnson, Volkov, \& Erwin}]{Johnson_2013}
Johnson, R.~E., Volkov, A.~N., \& Erwin, J.~T. 2013, The Astrophysical Journal,
  768, L4

\bibitem[{{Kausch} {et~al.}(2015){Kausch}, {Noll}, {Smette}, {Kimeswenger},
  {Barden}, {Szyszka}, {Jones}, {Sana}, {Horst}, \& {Kerber}}]{kausch15}
{Kausch}, W., {Noll}, S., {Smette}, A., {et~al.} 2015, \aap, 576, A78

\bibitem[{{Kempton} {et~al.}(2018){Kempton}, {Bean}, {Louie}, {Deming}, {Koll},
  {Mansfield}, {Christiansen}, {L{\'o}pez-Morales}, {Swain}, {Zellem},
  {Ballard}, {Barclay}, {Barstow}, {Batalha}, {Beatty}, {Berta-Thompson},
  {Birkby}, {Buchhave}, {Charbonneau}, {Cowan}, {Crossfield}, {de Val-Borro},
  {Doyon}, {Dragomir}, {Gaidos}, {Heng}, {Hu}, {Kane}, {Kreidberg}, {Mallonn},
  {Morley}, {Narita}, {Nascimbeni}, {Pall{\'e}}, {Quintana}, {Rauscher},
  {Seager}, {Shkolnik}, {Sing}, {Sozzetti}, {Stassun}, {Valenti}, \& {von
  Essen}}]{kempton18}
{Kempton}, E.~M.-R., {Bean}, J.~L., {Louie}, D.~R., {et~al.} 2018, \pasp, 130,
  114401

\bibitem[{{Kirk} {et~al.}(2020){Kirk}, {Alam}, {L{\'o}pez-Morales}, \&
  {Zeng}}]{kirk20}
{Kirk}, J., {Alam}, M.~K., {L{\'o}pez-Morales}, M., \& {Zeng}, L. 2020, \aj,
  159, 115

\bibitem[{{Kite} {et~al.}(2020){Kite}, {Fegley}, {Schaefer}, \&
  {Ford}}]{kite20}
{Kite}, E.~S., {Fegley}, Bruce, J., {Schaefer}, L., \& {Ford}, E.~B. 2020,
  \apj, 891, 111

\bibitem[{{Knutson} {et~al.}(2014){Knutson}, {Dragomir}, {Kreidberg},
  {Kempton}, {McCullough}, {Fortney}, {Bean}, {Gillon}, {Homeier}, \&
  {Howard}}]{knutson14}
{Knutson}, H.~A., {Dragomir}, D., {Kreidberg}, L., {et~al.} 2014, \apj, 794,
  155

\bibitem[{{Kreidberg} \& {Oklop{\v c}i{\'c}}(2018)}]{kreidberg18}
{Kreidberg}, L., \& {Oklop{\v c}i{\'c}}, A. 2018, Research Notes of the
  American Astronomical Society, 2, 44

\bibitem[{{Kreidberg} {et~al.}(2014){Kreidberg}, {Bean}, {D{\'e}sert},
  {Benneke}, {Deming}, {Stevenson}, {Seager}, {Berta-Thompson}, {Seifahrt}, \&
  {Homeier}}]{kreidberg14}
{Kreidberg}, L., {Bean}, J.~L., {D{\'e}sert}, J.-M., {et~al.} 2014, \nat, 505,
  69

\bibitem[{{Linsky} {et~al.}(2014){Linsky}, {Fontenla}, \&
  {France}}]{Linsky2014}
{Linsky}, J.~L., {Fontenla}, J., \& {France}, K. 2014, \apj, 780, 61

\bibitem[{{Louie} {et~al.}(2018){Louie}, {Deming}, {Albert}, {Bouma}, {Bean},
  \& {Lopez-Morales}}]{louie18}
{Louie}, D.~R., {Deming}, D., {Albert}, L., {et~al.} 2018, \pasp, 130, 044401

\bibitem[{{Loyd} {et~al.}(2016){Loyd}, {France}, {Youngblood}, {Schneider},
  {Brown}, {Hu}, {Linsky}, {Froning}, {Redfield}, {Rugheimer}, \&
  {Tian}}]{loyd16}
{Loyd}, R.~O.~P., {France}, K., {Youngblood}, A., {et~al.} 2016, \apj, 824, 102

\bibitem[{Malsky \& Rogers(2020)}]{Malsky_2020}
Malsky, I., \& Rogers, L.~A. 2020, The Astrophysical Journal, 896, 48

\bibitem[{{Mansfield} {et~al.}(2018){Mansfield}, {Bean}, {Oklop{\v c}i{\'c}},
  {Kreidberg}, {D{\'e}sert}, {Kempton}, {Line}, {Fortney}, {Henry}, {Mallonn},
  {Stevenson}, {Dragomir}, {Allart}, \& {Bourrier}}]{mansfield18}
{Mansfield}, M., {Bean}, J.~L., {Oklop{\v c}i{\'c}}, A., {et~al.} 2018, \apjl,
  868, L34

\bibitem[{{Martin} {et~al.}(2018){Martin}, {Fitzgerald}, {McLean}, {Doppmann},
  {Kassis}, {Aliado}, {Canfield}, {Johnson}, {Kress}, {Lanclos}, {Magnone},
  {Sohn}, {Wang}, \& {Weiss}}]{martin18}
{Martin}, E.~C., {Fitzgerald}, M.~P., {McLean}, I.~S., {et~al.} 2018, in
  Society of Photo-Optical Instrumentation Engineers (SPIE) Conference Series,
  Vol. 10702, \procspie, 107020A

\bibitem[{{McLean} {et~al.}(1998){McLean}, {Becklin}, {Bendiksen}, {Brims},
  {Canfield}, {Figer}, {Graham}, {Hare}, {Lacayanga}, {Larkin}, {Larson},
  {Levenson}, {Magnone}, {Teplitz}, \& {Wong}}]{mclean98}
{McLean}, I.~S., {Becklin}, E.~E., {Bendiksen}, O., {et~al.} 1998, Society of
  Photo-Optical Instrumentation Engineers (SPIE) Conference Series, Vol. 3354,
  {Design and development of NIRSPEC: a near-infrared echelle spectrograph for
  the Keck II telescope}, ed. A.~M. {Fowler}, 566--578

\bibitem[{{Miller} {et~al.}(2009){Miller}, {Fortney}, \&
  {Jackson}}]{2009ApJ...702.1413M}
{Miller}, N., {Fortney}, J.~J., \& {Jackson}, B. 2009, \apj, 702, 1413

\bibitem[{{Miller-Ricci} {et~al.}(2009){Miller-Ricci}, {Seager}, \&
  {Sasselov}}]{miller-ricci09}
{Miller-Ricci}, E., {Seager}, S., \& {Sasselov}, D. 2009, \apj, 690, 1056

\bibitem[{{Morley} {et~al.}(2015){Morley}, {Fortney}, {Marley}, {Zahnle},
  {Line}, {Kempton}, {Lewis}, \& {Cahoy}}]{morley15}
{Morley}, C.~V., {Fortney}, J.~J., {Marley}, M.~S., {et~al.} 2015, \apj, 815,
  110

\bibitem[{{Ninan} {et~al.}(2020){Ninan}, {Stefansson}, {Mahadevan}, {Bender},
  {Robertson}, {Ramsey}, {Terrien}, {Wright}, {Diddams}, {Kanodia}, {Cochran},
  {Endl}, {Ford}, {Fredrick}, {Halverson}, {Hearty}, {Jennings}, {Kaplan},
  {Lubar}, {Metcalf}, {Monson}, {Nitroy}, {Roy}, \& {Schwab}}]{ninan20}
{Ninan}, J.~P., {Stefansson}, G., {Mahadevan}, S., {et~al.} 2020, \apj, 894, 97

\bibitem[{{Niraula} {et~al.}(2017){Niraula}, {Redfield}, {Dai}, {Barrag{\'a}n},
  {Gandolfi}, {Cauley}, {Hirano}, {Korth}, {Smith}, {Prieto-Arranz}, {Grziwa},
  {Fridlund}, {Persson}, {Justesen}, {Winn}, {Albrecht}, {Cochran},
  {Csizmadia}, {Duvvuri}, {Endl}, {Hatzes}, {Livingston}, {Narita}, {Nespral},
  {Nowak}, {P{\"a}tzold}, {Palle}, \& {Van Eylen}}]{niraula17}
{Niraula}, P., {Redfield}, S., {Dai}, F., {et~al.} 2017, \aj, 154, 266

\bibitem[{{Nortmann} {et~al.}(2018){Nortmann}, {Pall{\'e}}, {Salz},
  {Sanz-Forcada}, {Nagel}, {Alonso-Floriano}, {Czesla}, {Yan}, {Chen},
  {Snellen}, {Zechmeister}, {Schmitt}, {L{\'o}pez-Puertas}, {Casasayas-Barris},
  {Bauer}, {Amado}, {Caballero}, {Dreizler}, {Henning}, {Lamp{\'o}n}, {Montes},
  {Molaverdikhani}, {Quirrenbach}, {Reiners}, {Ribas}, {S{\'a}nchez-L{\'o}pez},
  {Schneider}, \& {Zapatero Osorio}}]{nortmann18}
{Nortmann}, L., {Pall{\'e}}, E., {Salz}, M., {et~al.} 2018, Science, 362, 1388

\bibitem[{{Oklop{\v c}i{\'c}} \& {Hirata}(2018)}]{oklopcic18}
{Oklop{\v c}i{\'c}}, A., \& {Hirata}, C.~M. 2018, \apjl, 855, L11

\bibitem[{{Oklop{\v{c}}i{\'c}}(2019)}]{oklopcic19}
{Oklop{\v{c}}i{\'c}}, A. 2019, \apj, 881, 133

\bibitem[{{Otegi} {et~al.}(2020){Otegi}, {Bouchy}, \& {Helled}}]{otegi20}
{Otegi}, J.~F., {Bouchy}, F., \& {Helled}, R. 2020, \aap, 634, A43

\bibitem[{{Owen} \& {Wu}(2017)}]{owen17}
{Owen}, J.~E., \& {Wu}, Y. 2017, \apj, 847, 29

\bibitem[{{Palle} {et~al.}(2020){Palle}, {Nortmann}, {Casasayas-Barris},
  {Lamp{\'o}n}, {L{\'o}pez-Puertas}, {Caballero}, {Sanz-Forcada}, {Lara},
  {Nagel}, {Yan}, {Alonso-Floriano}, {Amado}, {Chen}, {Cifuentes},
  {Cort{\'e}s-Contreras}, {Czesla}, {Molaverdikhani}, {Montes}, {Passegger},
  {Quirrenbach}, {Reiners}, {Ribas}, {S{\'a}nchez-L{\'o}pez}, {Schweitzer},
  {Stangret}, {Zapatero Osorio}, \& {Zechmeister}}]{palle20}
{Palle}, E., {Nortmann}, L., {Casasayas-Barris}, N., {et~al.} 2020, \aap, 638,
  A61

\bibitem[{{Paxton} {et~al.}(2011){Paxton}, {Bildsten}, {Dotter}, {Herwig},
  {Lesaffre}, \& {Timmes}}]{Paxton2011}
{Paxton}, B., {Bildsten}, L., {Dotter}, A., {et~al.} 2011, \apjs, 192, 3

\bibitem[{{Paxton} {et~al.}(2013){Paxton}, {Cantiello}, {Arras}, {Bildsten},
  {Brown}, {Dotter}, {Mankovich}, {Montgomery}, {Stello}, {Timmes}, \&
  {Townsend}}]{Paxton2013}
{Paxton}, B., {Cantiello}, M., {Arras}, P., {et~al.} 2013, \apjs, 208, 4

\bibitem[{{Paxton} {et~al.}(2015){Paxton}, {Marchant}, {Schwab}, {Bauer},
  {Bildsten}, {Cantiello}, {Dessart}, {Farmer}, {Hu}, {Langer}, {Townsend},
  {Townsley}, \& {Timmes}}]{Paxton2015}
{Paxton}, B., {Marchant}, P., {Schwab}, J., {et~al.} 2015, \apjs, 220, 15

\bibitem[{{Paxton} {et~al.}(2018){Paxton}, {Schwab}, {Bauer}, {Bildsten},
  {Blinnikov}, {Duffell}, {Farmer}, {Goldberg}, {Marchant}, {Sorokina},
  {Thoul}, {Townsend}, \& {Timmes}}]{Paxton2018}
{Paxton}, B., {Schwab}, J., {Bauer}, E.~B., {et~al.} 2018, \apjs, 234, 34

\bibitem[{{Paxton} {et~al.}(2019){Paxton}, {Smolec}, {Schwab}, {Gautschy},
  {Bildsten}, {Cantiello}, {Dotter}, {Farmer}, {Goldberg}, {Jermyn}, {Kanbur},
  {Marchant}, {Thoul}, {Townsend}, {Wolf}, {Zhang}, \& {Timmes}}]{Paxton2019}
{Paxton}, B., {Smolec}, R., {Schwab}, J., {et~al.} 2019, \apjs, 243, 10

\bibitem[{{Prieto-Arranz} {et~al.}(2018){Prieto-Arranz}, {Palle}, {Gandolfi},
  {Barrag{\'a}n}, {Guenther}, {Dai}, {Fridlund}, {Hirano}, {Livingston},
  {Luque}, {Niraula}, {Persson}, {Redfield}, {Albrecht}, {Alonso},
  {Antoniciello}, {Cabrera}, {Cochran}, {Csizmadia}, {Deeg}, {Eigm{\"u}ller},
  {Endl}, {Erikson}, {Everett}, {Fukui}, {Grziwa}, {Hatzes}, {Hidalgo},
  {Hjorth}, {Korth}, {Lorenzo-Oliveira}, {Murgas}, {Narita}, {Nespral},
  {Nowak}, {P{\"a}tzold}, {Monta{\~n}ez Rodr{\'\i}guez}, {Rauer}, {Ribas},
  {Smith}, {Trifonov}, {Van Eylen}, \& {Winn}}]{prieto18}
{Prieto-Arranz}, J., {Palle}, E., {Gandolfi}, D., {et~al.} 2018, \aap, 618,
  A116

\bibitem[{{Rice} {et~al.}(2019){Rice}, {Malavolta}, {Mayo}, {Mortier},
  {Buchhave}, {Affer}, {Vanderburg}, {Lopez-Morales}, {Poretti}, {Zeng},
  {Cameron}, {Damasso}, {Coffinet}, {Latham}, {Bonomo}, {Bouchy},
  {Charbonneau}, {Dumusque}, {Figueira}, {Martinez Fiorenzano}, {Haywood},
  {Johnson}, {Lopez}, {Lovis}, {Mayor}, {Micela}, {Molinari}, {Nascimbeni},
  {Nava}, {Pepe}, {Phillips}, {Piotto}, {Sasselov}, {S{\'e}gransan},
  {Sozzetti}, {Udry}, \& {Watson}}]{rice19}
{Rice}, K., {Malavolta}, L., {Mayo}, A., {et~al.} 2019, \mnras,
  arXiv:1812.07302

\bibitem[{{Rodriguez} {et~al.}(2018){Rodriguez}, {Vanderburg}, {Eastman},
  {Mann}, {Crossfield}, {Ciardi}, {Latham}, \& {Quinn}}]{rodriguez18}
{Rodriguez}, J.~E., {Vanderburg}, A., {Eastman}, J.~D., {et~al.} 2018, \aj,
  155, 72

\bibitem[{{Rogers} {et~al.}(2011){Rogers}, {Bodenheimer}, {Lissauer}, \&
  {Seager}}]{rogers11}
{Rogers}, L.~A., {Bodenheimer}, P., {Lissauer}, J.~J., \& {Seager}, S. 2011,
  \apj, 738, 59

\bibitem[{{Rogers} \& {Seager}(2010{\natexlab{a}})}]{rogers10a}
{Rogers}, L.~A., \& {Seager}, S. 2010{\natexlab{a}}, \apj, 712, 974

\bibitem[{{Rogers} \& {Seager}(2010{\natexlab{b}})}]{rogers10b}
---. 2010{\natexlab{b}}, \apj, 716, 1208

\bibitem[{{Salz} {et~al.}(2016){Salz}, {Czesla}, {Schneider}, \&
  {Schmitt}}]{salz16}
{Salz}, M., {Czesla}, S., {Schneider}, P.~C., \& {Schmitt}, J.~H.~M.~M. 2016,
  \aap, 586, A75

\bibitem[{{Salz} {et~al.}(2018){Salz}, {Czesla}, {Schneider}, {Nagel},
  {Schmitt}, {Nortmann}, {Alonso-Floriano}, {L{\'o}pez-Puertas}, {Lamp{\'o}n},
  {Bauer}, {Snellen}, {Pall{\'e}}, {Caballero}, {Yan}, {Chen}, {Sanz-Forcada},
  {Amado}, {Quirrenbach}, {Ribas}, {Reiners}, {B{\'e}jar}, {Casasayas-Barris},
  {Cort{\'e}s-Contreras}, {Dreizler}, {Guenther}, {Henning}, {Jeffers},
  {Kaminski}, {K{\"u}rster}, {Lafarga}, {Lara}, {Molaverdikhani}, {Montes},
  {Morales}, {S{\'a}nchez-L{\'o}pez}, {Seifert}, {Zapatero Osorio}, \&
  {Zechmeister}}]{salz18}
{Salz}, M., {Czesla}, S., {Schneider}, P.~C., {et~al.} 2018, \aap, 620, A97

\bibitem[{{Sanz-Forcada} {et~al.}(2011){Sanz-Forcada}, {Micela}, {Ribas},
  {Pollock}, {Eiroa}, {Velasco}, {Solano}, \&
  {Garc{\'{\i}}a-{\'A}lvarez}}]{2011A&A...532A...6S}
{Sanz-Forcada}, J., {Micela}, G., {Ribas}, I., {et~al.} 2011, \aap, 532, A6

\bibitem[{{Seager} {et~al.}(2007){Seager}, {Kuchner}, {Hier-Majumder}, \&
  {Militzer}}]{seager07}
{Seager}, S., {Kuchner}, M., {Hier-Majumder}, C.~A., \& {Militzer}, B. 2007,
  \apj, 669, 1279

\bibitem[{{Smette} {et~al.}(2015){Smette}, {Sana}, {Noll}, {Horst}, {Kausch},
  {Kimeswenger}, {Barden}, {Szyszka}, {Jones}, {Gallenne}, {Vinther},
  {Ballester}, \& {Taylor}}]{smette15}
{Smette}, A., {Sana}, H., {Noll}, S., {et~al.} 2015, \aap, 576, A77

\bibitem[{{Southworth}(2011)}]{southworth11}
{Southworth}, J. 2011, \mnras, 417, 2166

\bibitem[{{Spake} {et~al.}(2018){Spake}, {Sing}, {Evans}, {Oklop{\v c}i{\'c}},
  {Bourrier}, {Kreidberg}, {Rackham}, {Irwin}, {Ehrenreich}, {Wyttenbach},
  {Wakeford}, {Zhou}, {Chubb}, {Nikolov}, {Goyal}, {Henry}, {Williamson},
  {Blumenthal}, {Anderson}, {Hellier}, {Charbonneau}, {Udry}, \&
  {Madhusudhan}}]{spake18}
{Spake}, J.~J., {Sing}, D.~K., {Evans}, T.~M., {et~al.} 2018, \nat, 557, 68

\bibitem[{{Sperauskas} {et~al.}(2016){Sperauskas},
  {Barta{\v{s}}i{\={u}}t{\.{e}}}, {Boyle}, {Deveikis}, {Raudeli{\={u}}nas}, \&
  {Upgren}}]{sperauskas16}
{Sperauskas}, J., {Barta{\v{s}}i{\={u}}t{\.{e}}}, S., {Boyle}, R.~P., {et~al.}
  2016, \aap, 596, A116

\bibitem[{{Teske} {et~al.}(2018){Teske}, {Wang}, {Wolfgang}, {Dai}, {Shectman},
  {Butler}, {Crane}, \& {Thompson}}]{teske18}
{Teske}, J.~K., {Wang}, S., {Wolfgang}, A., {et~al.} 2018, \aj, 155, 148

\bibitem[{{Tsiaras} {et~al.}(2019){Tsiaras}, {Waldmann}, {Tinetti}, {Tennyson},
  \& {Yurchenko}}]{tsiaras19}
{Tsiaras}, A., {Waldmann}, I.~P., {Tinetti}, G., {Tennyson}, J., \&
  {Yurchenko}, S.~N. 2019, Nature Astronomy, 3, 1086

\bibitem[{{Van Eylen} {et~al.}(2018){Van Eylen}, {Agentoft}, {Lundkvist},
  {Kjeldsen}, {Owen}, {Fulton}, {Petigura}, \& {Snellen}}]{vaneylen18}
{Van Eylen}, V., {Agentoft}, C., {Lundkvist}, M.~S., {et~al.} 2018, \mnras,
  479, 4786

\bibitem[{{Van Grootel} {et~al.}(2014){Van Grootel}, {Gillon}, {Valencia},
  {Madhusudhan}, {Dragomir}, {Howe}, {Burrows}, {Demory}, {Deming},
  {Ehrenreich}, {Lovis}, {Mayor}, {Pepe}, {Queloz}, {Scuflaire}, {Seager},
  {Segransan}, \& {Udry}}]{VanGrootel2014}
{Van Grootel}, V., {Gillon}, M., {Valencia}, D., {et~al.} 2014, \apj, 786, 2

\bibitem[{{Vissapragada} {et~al.}(2020){Vissapragada}, {Knutson}, {Jovanovic},
  {Harada}, {Oklop{\v{c}}i{\'c}}, {Eriksen}, {Mawet}, {Millar-Blanchaer},
  {Tinyanont}, \& {Vasisht}}]{vissa20}
{Vissapragada}, S., {Knutson}, H.~A., {Jovanovic}, N., {et~al.} 2020, \aj, 159,
  278

\bibitem[{{Youngblood} {et~al.}(2016){Youngblood}, {France}, {Loyd}, {Linsky},
  {Redfield}, {Schneider}, {Wood}, {Brown}, {Froning}, {Miguel}, {Rugheimer},
  \& {Walkowicz}}]{youngblood16}
{Youngblood}, A., {France}, K., {Loyd}, R.~O.~P., {et~al.} 2016, \apj, 824, 101

\bibitem[{{Youngblood} {et~al.}(2019){Youngblood}, {France}, {Koskinen},
  {Fossati}, {Amerstorfer}, {Lichtenegger}, {Drake}, {Mason}, {Fleming},
  {Allred}, {Berta-Thompson}, {Bourrier}, {Froning}, {Garraffo}, {Gronoff},
  {Jin}, {Kowalski}, \& {Osten}}]{Youngblood2019}
{Youngblood}, A., {France}, K., {Koskinen}, T., {et~al.} 2019, \baas, 51, 320

\end{thebibliography}
\end{document}